\documentclass[10pt]{article}
\usepackage[centering]{geometry}

\usepackage{amsmath,amssymb,mathrsfs}
\usepackage{bbm} 

\usepackage[dvips]{graphicx}

\usepackage{pstricks}

\usepackage[dvips,bookmarks=true]{hyperref} 
\hypersetup{pdfstartview=FitH,pdfhighlight=/O,colorlinks=false}

\newcommand{\be}{\begin{equation}}
\newcommand{\ee}{\end{equation}}
\newcommand{\ben}{\begin{equation*}}
\newcommand{\een}{\end{equation*}}

\newcommand{\mc}[1]{\mathcal{#1}}
\newcommand{\ms}[1]{\mathscr{#1}}
\newcommand{\mbb}[1]{\mathbb{#1}}

\newcommand{\id}{\mathbbm{1}}

\newcommand{\de}{\delta}
\newcommand{\p}{\partial}

\newcommand{\eps}{\varepsilon}

\newcommand{\Tr}{{\rm Tr}}

\newcommand{\mdots}{,.\,.\,,}

\title{\Large{\textbf{The perturbative Regge-calculus regime}}\\
 \Large{\textbf{of Loop Quantum Gravity}}}

\author{Eugenio Bianchi\footnote{\texttt{e.bianchi@sns.it}}\\[.35em]
\small{Scuola  Normale Superiore \& INFN Pisa,
Piazza dei Cavalieri 7, I-56126 Pisa,  Italy }
}
\author{Eugenio Bianchi${}^{1}$\footnote{\texttt{e.bianchi@sns.it}}
$\;$ and Leonardo Modesto${}^{2,3}$\footnote{\texttt{modesto@bo.infn.it}}\\[.35em]
\small{${}^1$Scuola  Normale Superiore \& INFN Pisa,
Piazza dei Cavalieri 7, I-56126 Pisa,  Italy }\\
\small{${}^2$Department of Physics, Bologna University \& INFN Bologna, V. Irnerio 46, I-40126 Bologna, Italy}\\
\small{${}^3$Centre de Physique Th\'eorique de Luminy, Universit\'e de la M\'editerran\'ee, F-13288 Marseille, France}
}

\date{\small \today}

\begin{document}

\maketitle

\begin{abstract}
The relation between Loop Quantum Gravity and Regge calculus has been pointed out many times in the literature. In particular the large spin asymptotics of the Barrett-Crane vertex amplitude is known to be related to the Regge action. In this paper we study a semiclassical regime of Loop Quantum Gravity and show that it admits an effective description in terms of perturbative area-Regge-calculus. The regime of interest is identified by a class of states given by superpositions of four-valent spin networks, peaked on large spins. 
 
As a probe of the dynamics in this regime, we compute explicitly two- and three-area correlation functions at the vertex amplitude level. We find that they match with the ones computed perturbatively in area-Regge-calculus with a single 4-simplex, once a specific perturbative action and measure have been chosen in the Regge-calculus path integral. 
Correlations of other geometric operators and the existence of this regime for other models for the dynamics are briefly discussed.

\begin{flushleft}
\end{flushleft}
\end{abstract}


\section{Introduction}\label{sec:intro}
The relation between Loop Quantum Gravity \cite{Rovelli:2004tv,Ashtekar:2004eh,Thiemann:book2007} and Regge calculus \cite{Regge:1961px} has been pointed out many times in the literature, both at the kinematical and at the dynamical level. 

The fact that in $SU(2)$ representation theory some inequalities and symmetries appear which have a geometrical interpretation in terms of flat $n$-simplices is known since long ago \cite{Wigner:1959,Landau:1958III,PonzanoRegge:1968}. Its relevance here comes from the key role played by $SU(2)$ representation theory in Loop Quantum Gravity \cite{DePietri:1996pj} on the one side, and by flat $n$-simplices in the Regge approximation to General Relativity \cite{MTW:1973} on the other. For instance, restricting attention to kinematics, four-valent nodes of a $SU(2)$-spin-network share some geometrical properties with flat tetrahedra and can be considered their quantum version \cite{Immirzi:1996dr,Barbieri:1997ks,Baez:1999tk}. Moreover, appropriate superpositions of these `quantum tetrahedron'-states can be shown to describe exactly the geometry of a classical tetrahedron \cite{Rovelli:2006fw}. 

Spin foam models \cite{Reisenberger:1994aw,Baez:1997zt,Perez:2003vx} provide a covariant way to code the dynamics within the framework of Loop Quantum Gravity. Despite the fact that the spin foam formalism can be motivated from various independent perspectives, building a model always requires some input. For instance, this input can come under the form of a classical action principle and a procedure to make sense of its formal path integral. In any case, once a model is available, one can forget the heuristics and start studying its properties. In order for a specific model to be claimed to be a model for quantum gravity (and not to be any other diffeomorphism invariant theory) some check of its semiclassical properties is necessary.  

The most well-studied model for quantum gravity is certainly the one proposed by Barrett and Crane in \cite{Barrett:1997gw}. It is given by a sum over intertwiners of the $\{15j\}$ symbol squared. More recently Engle, Pereira and Rovelli \cite{Engle:2007uq,Engle:2007qf} proposed an improvement from the Loop Quantum Gravity point of view of the Barrett-Crane model. Both models take as starting point some form of classical simplicial gravity and end up with an elementary ``symbol'' in $SU(2)$ representation theory. However this by itself is not enough to guarantee that the dynamics admits Regge gravity as a semiclassical regime. An indication that in fact this is the case, at least for the Barrett-Crane model, was presented in \cite{Barrett:1998gs} where it was shown that the Barrett-Crane vertex amplitude has a large spin asymptotics of the form
\be\label{eq:BCasymptotics}
A_{\textrm{BC}}(j_{mn})\sim P(j_{mn}) \cos\big( S_{\textrm{Regge}}(j_{mn}) + \frac{\pi}{4}\,\big) + D(j_{mn})
\ee
where $S_{\textrm{Regge}}$ is the Regge action for a single $4$-simplex with faces having areas related to the spins $j_{mn}$. This result is analogous to Ponzano and Regge asymptotic formula for the Wigner $\{6j\}$ symbol giving the cosine of Regge action in three dimensions \cite{PonzanoRegge:1968}. It can be considered as a formal semiclassical limit, with $\hbar\to 0$ and the areas kept fixed. 

Expression (\ref{eq:BCasymptotics}) was obtained starting from an integral formula for the Barrett-Crane vertex \cite{Barrett:1998at} and using a stationary phase approximation. In \cite{Barrett:1998gs}, Barrett and Williams shown that the stationarity condition for the phase reproduces exactly the form of the Schl\"afli differential identity satisfied by the Regge action (i.e. the discrete version of Palatini identity). This identifies the Regge action up to an overall scale. 
Among the stationary points there are both ones corresponding to four dimensional non-degenerate simplicial geometries and ones corresponding to low-dimensional degenerate and non-degenerate geometries. The latter ones contribute to the $D(j_{mn})$ term in (\ref{eq:BCasymptotics}). Numerical computations \cite{Baez:2002rx} and analytical calculations \cite{Barrett:2002ur,Freidel:2002mj} show that in fact the degenerate configurations give the dominant contribution to the asymptotic formula. This has raised doubts on the physical viability of the model. 

However it should be noticed that simply taking the large spin limit is not enough to identify \emph{a} semiclassical regime, exactly as taking the large distance limit in the Hydrogen atom transition amplitude kernel $\langle r_2 |U(T)|r_1\rangle$ is not enough to identify its semiclassical regime. What is needed is a state peaked at large distance from the attractive center and on \emph{a} specific momentum. The state codes the initial conditions for a classical orbit and the dynamics of this state identifies a Keplerian semiclassical regime as can be shown computing correlations of position at different times. Taking into account this fact, in \cite{Rovelli:2005yj,Bianchi:2006uf} a study of the dynamics of the Barrett-Crane model on a boundary semiclassical state peaked on a specific intrinsic and extrinsic $3$-geometry was presented and the following general mechanism was shown to be at work: of the right-hand side of (\ref{eq:BCasymptotics}) only the exponential of $i$ times the Regge action contributes to the two-spin correlation function, while the other terms are suppressed by the phase of the semiclassical state. In this paper we strengthen and extend this result. \\

The main initial motivation for the present work was to compute the three-spin correlation function at leading order in the framework described above. A task of this kind requires some difficulties to be overcome the first. In \cite{Rovelli:2005yj,Bianchi:2006uf}, in order to compute the spin-spin correlation function needed for the analysis of the graviton propagator,  formula (\ref{eq:BCasymptotics}) was widely used. However the detailed dependence of $D(j_{mn})$ and $P(j_{mn})$ on the fluctuation of a large spin $j_0$ was not given in \cite{Barrett:2002ur,Freidel:2002mj} and it cannot be deduced from numerical computations \cite{Baez:2002rx}. In \cite{Rovelli:2005yj,Bianchi:2006uf} $D(j_{mn})$ was assumed to be slowly varying with the fluctuation $\de j_{mn}=j_{mn}-j_0$ and $P(j_{mn})$ was assumed to be independent from the fluctuation $\de j_{mn}$, at least at leading order in $j_0^{-1}$. Moreover, the same function $S_{\textrm{Regge}}(j_{mn})$ is not completely under control as it is given as a function of the spins which are related to the area of the faces of the $4$-simplex \cite{Barrett:1997tx,Makela:1998wi,Makela:2000ej} and not to its edge lengths: besides the fact that 

\vspace{-.8em}

\begin{center}
- \begin{minipage}[t]{.95 \textwidth}
for a single $4$-simplex there are isolated configurations for which giving the areas does not completely fix the geometry of the $4$-simplex, and
\end{minipage}

\vspace{.2em}

- \begin{minipage}[t]{.95 \textwidth}
for more than one $4$-simplex some constraints between the areas arise which are still to be worked out, 
\end{minipage}
\end{center}

\vspace{-.7em}


\noindent the main problem is that an explicit expression for the deficit angles in terms of the areas is still not available. Nevertheless, the relation between area-Regge-calculus and Loop Quantum Gravity pointed out by Rovelli in \cite{Rovelli:1993kc} seems to be both inevitable and desirable, as acknowledged in \cite{Regge:2000wu}. The class of difficulties discussed above can be completely avoided if one manages to compute the  spin-spin correlation function directly from the Barrett-Crane vertex formula, without using its asymptotic expansion (\ref{eq:BCasymptotics}). This was done in \cite{Livine:2006it} where a suited boundary semiclassical state was introduced in order to perform analytically the calculation\footnote{For a similar analysis in three dimensions, see \cite{Speziale:2005ma,Livine:2006ab}.}. The result confirms the scaling found in \cite{Rovelli:2005yj,Bianchi:2006uf} and puts it on stronger grounds. However the analysis in \cite{Livine:2006it} presents two disadvantages: (i) it hides completely the relation with area-Regge-calculus at any intermediate step by construction and (ii) the new semiclassical state is not suited for a phenomenological parametrization beyond the leading order.

The route we follow here is to separate the two problems. As in \cite{Livine:2006it}, we study correlation functions in Loop Quantum Gravity without using expression (\ref{eq:BCasymptotics}), but still using the semiclassical state used in \cite{Rovelli:2005yj,Bianchi:2006uf} which allows a phenomenological parametrization for corrections contributing to the three-spin correlation function. This is possible thanks to the two following results:
\begin{itemize}
\item writing the Barrett-Crane vertex as an integral over the group manifold $(S^3)^5$, we identify the region of the manifold which gives the dominant contribution to the $n$-spin correlation functions. We find that it is given by a bounded domain $\mc{B}$ of radius of order $1/\sqrt{j_0}$. The integral on its complement gives an exponentially suppressed contribution;
\item the dominant contribution can be computed explicitly thanks to the properties of the two parameters coming into play, the large mean spin $j_0$ and the small relative fluctuation $k_{mn}=\de j_{mn}/j_0$, as coded by the boundary semiclassical state. Hence we have both

\hspace{1em} - an asymptotic expansion in $j_0^{-1}$ and

\hspace{1em} - a perturbative expansion in $k_{mn}$

of the Barrett-Crane integral restricted to the domain $\mc{B}$.
\end{itemize}
These results allow to compute again, with a different technique, the two-spin correlation function computed in \cite{Rovelli:2005yj,Bianchi:2006uf,Livine:2006it}. Moreover, using this technique, we compute the three-spin correlation function. Its relevance for the study of the semiclassical regime of Loop Quantum Gravity will be discussed in a separate paper \cite{BMR:toappear}. Here it is enough to notice that, as the two-spin correlation function is related to the graviton propagator, the three-spin correlation function is related to the $3$-graviton Feynman-interaction-vertex. But there is much more: the analysis discussed above identifies a perturbative regime within the non-perturbative Loop Quantum Gravity framework. Thanks to the boundary semiclassical state, the dynamics can be described perturbatively in the fluctuation $\de j_{mn}$ in terms of a perturbative action and a perturbative measure.

On the other hand we study \emph{perturbative} quantum area-Regge-calculus and compute two- and three-area correlation functions for triangles belonging to a single $4$-simplex. By perturbative we mean that we fix a background Regge skeleton and background values of the areas so that they describe flat space. Then we quantize perturbatively the fluctuations of the areas in a path-integral fashion. The quantum theory is defined in terms of a perturbative action and a perturbative measure. This approach was developed in \cite{Rocek:1982fr,Rocek:1982tj} for length-Regge-calculus. Starting with a single regular $4$-simplex as background, area variables pose no problem as the Jacobian from lengths to areas is well-defined. While the perturbative action can be obtained from the standard Regge action, the measure in the integral over area-fluctuations has to be postulated. Many proposals have appeared in the literature for the non-perturbative measure over lengths and, although the specific measure does not affect area correlation functions at leading order, it does at higher orders.  

This approach of keeping Loop Quantum Gravity with semiclassical states on the one side and perturbative quantum Regge calculus on the other allows a direct comparison between the two frameworks. We find that a complete matching can be achieved once a specific non-trivial perturbative measure has been chosen on the Regge calculus side. 

The paper is written in a self-contained form and is organized as follows. Sections \ref{sec:large-scale}-\ref{sec:conclusion LQG} are dedicated to the Loop Quantum Gravity calculation. In particular in section \ref{sec:large-scale} we introduce the framework and discuss our assumptions; in section \ref{sec:integral formulae} we give a representation of the Barrett-Crane vertex as an integral over $(S^3)^5$; in section \ref{sec:dominant contribution} we identify the region of $(S^3)^5$ in the integral  which gives the dominant contribution to correlations and in section \ref{sec:stationary phase} we compute such contribution using an asymptotic expansion. We conclude the Loop Quantum Gravity analysis in section \ref{sec:conclusion LQG} where we give an effective description of the Barrett-Crane model in terms of a perturbative action and a perturbative measure. Moreover, in section \ref{sec:conclusion LQG}, we give explicit expressions of two-area and three-area correlation functions on the chosen semiclassical state.
Section \ref{sec:intro regge calculus} is dedicated to the perturbative quantum Regge calculus calculations. We set the framework for the calculation of area correlation functions and give an explicit perturbative expression for the area-Regge-calculus action. The comparison between the Loop Quantum Gravity calculation and the Regge calculus one is presented in section \ref{sec:comparison LQG RC} which concludes our analysis. The last section of the paper, section \ref{sec:discussion}, is dedicated to a discussion of the result, to an analysis of some perspectives it opens and to the possibility of using the techniques introduced here to test the recently proposed new models for the dynamics \cite{Engle:2007uq,Engle:2007qf,Livine:2007vk,Livine:2007ya,Freidel:2007py,Alexandrov:2007pq}.

\section{Large-scale correlations in spinfoam models}\label{sec:large-scale}
In this section we introduce the setting for the calculation of correlation functions in Loop Quantum Gravity. The boundary amplitude formalism is widely used both here in the Loop Quantum Gravity calculations and in section \ref{sec:intro regge calculus} in the Regge calculus calculation.

In the boundary amplitude formalism one considers regions $\mc{R}$ of the $4$-manifold $\mc{M}$ and assigns (i) a Hilbert space of states $\mc{H}_{\Sigma}$ to the boundary manifold $\Sigma=\p\mc{R}$, and (ii) a map $W_{\mc{R}}:\mc{H}_{\Sigma}\to \mbb{C}$ which codifies the dynamics in the region $\mc{R}$. This approach has been advocated in various forms in \cite{Rovelli:2004tv}, in \cite{Oeckl:2003pw,Oeckl:2003vu,Oeckl:2005bv,Oeckl:2005bw}, in \cite{Hayward:1992ix} and in \cite{Segal:1987sk,Atiyah:1989vu,Segal:1998itp}. This approach presents both technical and conceptual advantages. On the technical side, it avoids the need to prescribe the semiclassical asymptotic behaviour at spatial infinity \cite{Conrady:2003en}. In a sense, it could be the route to properly define asymptotic flatness in quantum gravity. On the conceptual side, it shifts the attention from ``quantum cosmology'' \cite{Hartle:1983ai}  to quantum gravity and the hope is that it could open the possibility to calculate scattering amplitudes in background independent approaches \cite{Modesto:2005sj}.

\subsection{The boundary amplitude formalism at work}

Here we shall take a minimalist approach (the reader can refer to \cite{Bianchi:2006uf} for the general philosophy). We consider $4$-dimensional Riemannian Loop Quantum Gravity with the dynamics implemented using the spinfoam formalism. In particular we study the Barrett-Crane spinfoam model \cite{Barrett:1997gw} and some of its modifications \cite{Baez:2002aw}, \cite{Engle:2007uq}. We restrict attention to a subspace of the boundary Hilbert space, the space of states having graphs containing only $4$-valent nodes. The Barrett-Crane model is well defined on this subspace. 

As well-known \cite{Rovelli:2004tv}, in Loop Quantum Gravity the Hamiltonian constraint acts on a spin-network state non-trivially only at nodes. In spinfoam models, the action of the Hamiltonian constraint at a node is given in terms of a spinfoam vertex amplitude \cite{Reisenberger:1996pu}. In particular, the Hamiltonian constraint corresponding to the Barrett-Crane spinfoam model is supposed to act in the following way on a state with a $4$-valent node

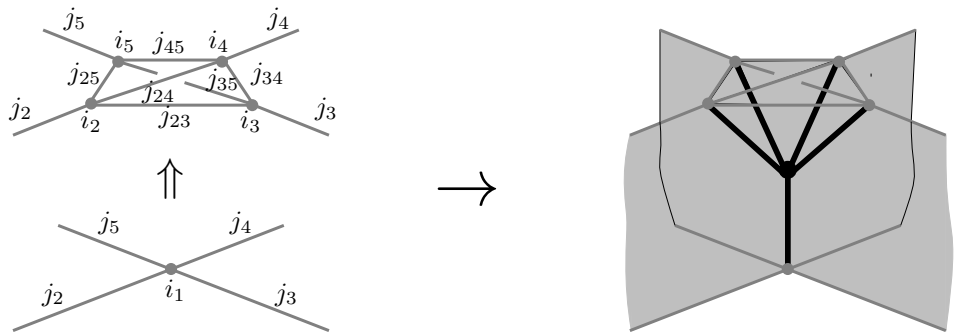
\begin{minipage}[t]{\textwidth}
\begin{minipage}[b]{0.38\textwidth}
\begin{pspicture}(0,-2.2992187)(4.8828125,2.3192186)
\psline[linewidth=0.04cm,linecolor=gray](0.2609375,-2.2792187)(3.8609376,-0.87921876)
\psline[linewidth=0.04cm,linecolor=gray](0.8609375,-0.87921876)(4.4609375,-2.2792187)
\psline[linewidth=0.04cm,linecolor=gray](0.2609375,0.32078126)(1.2609375,0.72078127)
\psline[linewidth=0.04cm,linecolor=gray](4.4609375,0.32078126)(3.4609375,0.72078127)
\psline[linewidth=0.04cm,linecolor=gray](4.0609374,1.7207812)(3.0609374,1.3207812)
\psline[linewidth=0.04cm,linecolor=gray](0.6609375,1.7207812)(1.6609375,1.3207812)
\psline[linewidth=0.04cm,linecolor=gray](1.6609375,1.3207812)(3.0609374,1.3207812)
\psline[linewidth=0.04cm,linecolor=gray](1.2609375,0.72078127)(3.4609375,0.72078127)
\psline[linewidth=0.04cm,linecolor=gray](1.2609375,0.72078127)(1.6609375,1.3207812)
\psline[linewidth=0.04cm,linecolor=gray](3.0609374,1.3207812)(3.4609375,0.72078127)
\psline[linewidth=0.04cm,linecolor=gray](1.2609375,0.72078127)(3.0609374,1.3207812)
\psline[linewidth=0.04cm,linecolor=gray](1.6609375,1.3007812)(2.1809375,1.1407813)
\psline[linewidth=0.04cm,linecolor=gray](2.5409374,1.0207813)(3.4209375,0.74078125)
\psdots[dotsize=0.16,linecolor=gray](2.3609376,-1.4592187)
\psdots[dotsize=0.16,linecolor=gray](1.3009375,0.74078125)
\psdots[dotsize=0.16,linecolor=gray](3.4409375,0.72078127)
\psdots[dotsize=0.16,linecolor=gray](3.0409374,1.3007812)
\psdots[dotsize=0.16,linecolor=gray](1.6609375,1.3007812)
\rput(.8,-1.8){$j_2$}
\rput(3.9,-1.8){$j_3$}
\rput(3.3,-0.85){$j_4$}
\rput(1.55,-0.85){$j_5$}
\rput(2.42,-1.75){$i_1$}
\rput(2.35,-0.25){\LARGE $\Uparrow$}
\rput(0.37,0.67){$j_2$}
\rput(4.43,0.67){$j_3$}
\rput(3.8,1.87){$j_4$}
\rput(1.09,1.87){$j_5$}
\rput(1.30,0.50){$i_2$}
\rput(3.4223437,0.50){$i_3$}
\rput(3.0,1.57){$i_4$}
\rput(1.75,1.57){$i_5$}
\rput(2.42,0.57){$j_{23}$}
\rput(2.2,0.9){$j_{24}$}
\rput(1.22,1.13){$j_{25}$}
\rput(3.65,1.13){$j_{34}$}
\rput(3.05,1.05){$j_{35}$}
\rput(2.3423438,1.55){$j_{45}$}
\end{pspicture} 
\end{minipage}
\begin{minipage}[b]{0.13\textwidth}
{\Huge $\rightarrow$}\\
${}$\\
${}$\\
${}$\\
${}$
\end{minipage}
\begin{minipage}[b]{0.45\textwidth}
\begin{pspicture}(0,-2.02)(4,2.02)
\pspolygon[linewidth=0.0020,fillstyle=solid,fillcolor=lightgray](3.26,1.6)(2.12,1.2)(1.96,1.58)
\pspolygon[linewidth=0.0020,fillstyle=solid,fillcolor=lightgray](3.3,1.58)(3.02,1.0)(3.72,1.0)
\pspolygon[linewidth=0.0020,fillstyle=solid,fillcolor=lightgray](2.1,1.2)(1.54,1.0)(1.92,1.56)
\pspolygon[linewidth=0.0020,fillstyle=solid,fillcolor=lightgray](4.32,2.0)(4.26,0.78)(3.72,1.0)(3.32,1.58)(3.48,1.66)
\pspolygon[linewidth=0.0020,fillstyle=solid,fillcolor=lightgray](0.92,2.0)(0.92,0.76)(1.5,0.98)(1.92,1.6)(1.82,1.64)
\pspolygon[linewidth=0.0020,fillstyle=solid,fillcolor=lightgray](1.56,1.02)(3.26,1.58)(3.04,1.0)(2.86,1.0)
\pspolygon[linewidth=0.0020,fillstyle=solid,fillcolor=lightgray](1.58,1.0)(3.66,1.0)(2.62,0.08)
\psbezier[linewidth=0.0020,linecolor=lightgray,fillstyle=solid,fillcolor=lightgray](4.72,0.6)(4.7,-1.0)(4.94,-0.32)(4.72,-2.0)
\pspolygon[linewidth=0.0020,linecolor=lightgray,fillstyle=solid,fillcolor=lightgray](4.72,-2.0)(2.62,-1.18)(2.64,0.08)(3.7,1.0)(4.72,0.6)(4.72,-0.76)
\psbezier[linewidth=0.0020,linecolor=lightgray,fillstyle=solid,fillcolor=lightgray](0.52,0.6)(0.38,0.02)(0.4775044,-0.31918114)(0.48,-0.56)(0.4824956,-0.80081886)(0.42,-1.32)(0.54,-2.0)
\pspolygon[linewidth=0.0020,linecolor=lightgray,fillstyle=solid,fillcolor=lightgray](0.52,0.6)(1.52,1.0)(2.6,0.06)(2.62,-1.16)(0.54,-2.0)(0.52,-0.84)
\psdots[dotsize=0.24](2.62,0.14)
\psline[linewidth=0.076cm](1.92,1.58)(2.6,0.1)
\psline[linewidth=0.076cm](1.54,1.02)(2.6,0.06)
\psline[linewidth=0.076cm](3.28,1.58)(2.6,0.06)
\psline[linewidth=0.076cm](3.68,0.98)(2.6,0.04)
\psline[linewidth=0.076cm](2.62,0.08)(2.62,-1.16)
\psline[linewidth=0.04cm,linecolor=gray](0.52,-2.0)(4.12,-0.6)
\psline[linewidth=0.04cm,linecolor=gray](1.12,-0.6)(4.72,-2.0)
\psline[linewidth=0.04cm,linecolor=gray](0.52,0.6)(1.52,1.0)
\psline[linewidth=0.04cm,linecolor=gray](4.72,0.6)(3.72,1.0)
\psline[linewidth=0.04cm,linecolor=gray](4.32,2.0)(3.32,1.6)
\psline[linewidth=0.04cm,linecolor=gray](0.92,2.0)(1.92,1.6)
\psline[linewidth=0.04cm,linecolor=gray](1.92,1.6)(3.32,1.6)
\psline[linewidth=0.04cm,linecolor=gray](1.52,1.0)(3.72,1.0)
\psline[linewidth=0.04cm](3.72,1.4)(3.72,1.4)
\psline[linewidth=0.04cm,linecolor=gray](1.52,1.0)(1.92,1.6)
\psline[linewidth=0.04cm,linecolor=gray](3.32,1.6)(3.72,1.0)
\psline[linewidth=0.04cm,linecolor=gray](1.52,1.0)(3.32,1.6)
\psline[linewidth=0.04cm,linecolor=gray](1.92,1.58)(2.44,1.42)
\psline[linewidth=0.04cm,linecolor=gray](2.8,1.3)(3.68,1.02)
\psdots[dotsize=0.16,linecolor=gray](2.62,-1.18)
\psdots[dotsize=0.16,linecolor=gray](1.56,1.02)
\psdots[dotsize=0.16,linecolor=gray](3.7,1.0)
\psdots[dotsize=0.16,linecolor=gray](3.3,1.58)
\psdots[dotsize=0.16,linecolor=gray](1.92,1.58)
\psbezier[linewidth=0.0020,fillstyle=solid,fillcolor=lightgray](0.92,0.78)(0.98,0.04)(0.78,0.48)(1.12,-0.6)
\psbezier[linewidth=0.0020,fillstyle=solid,fillcolor=lightgray](4.26,0.78)(4.26,-0.02)(4.42,0.24)(4.12,-0.6)
\end{pspicture} 
\end{minipage}
\end{minipage}

\vspace{1.5em}

\noindent where on the right-hand side we have depicted the spinfoam corresponding to it. Such transition amplitude can be written in the following way:
\be\label{eq:<4n|H|1n>}
\langle 
\begin{pspicture}(0,-0.10)(0.7,0.27)
\scalebox{1} 
{
\psline[linewidth=0.025cm](0.16,-0.11)(0.0,-0.25)
\psline[linewidth=0.025cm](0.44,-0.11)(0.6,-0.25)
\psline[linewidth=0.025cm](0.6,0.25)(0.44,0.13)
\psline[linewidth=0.025cm](0.0,0.25)(0.16,0.13)
\psline[linewidth=0.025cm](0.16,0.13)(0.44,-0.11)
\psline[linewidth=0.025cm](0.44,0.13)(0.16,0.13)
\psline[linewidth=0.025cm](0.16,-0.11)(0.16,0.13)
\psline[linewidth=0.025cm](0.16,-0.11)(0.44,-0.11)
\psline[linewidth=0.025cm](0.44,0.13)(0.44,-0.11)
\psline[linewidth=0.025cm](0.16,-0.11)(0.26,-0.03)
\psline[linewidth=0.025cm](0.44,0.13)(0.34,0.05)
\psdots[dotsize=0.1](0.16,0.13)
\psdots[dotsize=0.1](0.44,0.13)
\psdots[dotsize=0.1](0.44,-0.11)
\psdots[dotsize=0.1](0.16,-0.11)
}
\end{pspicture}
| \hat{H}|
\begin{pspicture}(0,-0.10)(0.7,0.27)
\scalebox{1} 
{
\psline[linewidth=0.025cm](0.0,0.25)(0.6,-0.25)
\psline[linewidth=0.025cm](0.6,0.25)(0.0,-0.25)
\psdots[dotsize=0.1](0.3,0.0)
}
\end{pspicture}
\rangle=\big(\prod_f A_f(j_l)\big)A_v(j_l,i_n)\;.
\ee

\vspace{-.5 em}

\noindent To capture the role of the vertex amplitude, this formula is best written in the boundary amplitude formalism introduced above. The strategy is the following: (i) cut out from the two-complex a $4$-ball $B_4$ containing a spinfoam vertex $v$ as shown in the following picture

\vspace{1 em} 

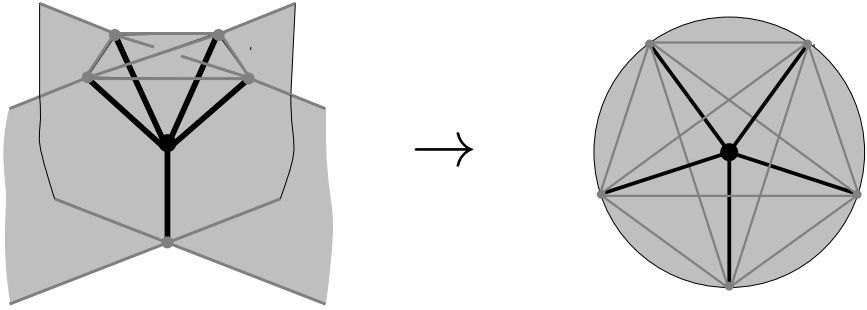
\begin{minipage}[c]{\textwidth}
\begin{minipage}[c]{0.38\textwidth}
\begin{pspicture}(0,-2.02)(4.941,2.02)
\pspolygon[linewidth=0.0020,fillstyle=solid,fillcolor=lightgray](3.26,1.6)(2.12,1.2)(1.96,1.58)
\pspolygon[linewidth=0.0020,fillstyle=solid,fillcolor=lightgray](3.3,1.58)(3.02,1.0)(3.72,1.0)
\pspolygon[linewidth=0.0020,fillstyle=solid,fillcolor=lightgray](2.1,1.2)(1.54,1.0)(1.92,1.56)
\pspolygon[linewidth=0.0020,fillstyle=solid,fillcolor=lightgray](4.32,2.0)(4.26,0.78)(3.72,1.0)(3.32,1.58)(3.48,1.66)
\pspolygon[linewidth=0.0020,fillstyle=solid,fillcolor=lightgray](0.92,2.0)(0.92,0.76)(1.5,0.98)(1.92,1.6)(1.82,1.64)
\pspolygon[linewidth=0.0020,fillstyle=solid,fillcolor=lightgray](1.56,1.02)(3.26,1.58)(3.04,1.0)(2.86,1.0)
\pspolygon[linewidth=0.0020,fillstyle=solid,fillcolor=lightgray](1.58,1.0)(3.66,1.0)(2.62,0.08)
\psbezier[linewidth=0.0020,linecolor=lightgray,fillstyle=solid,fillcolor=lightgray](4.72,0.6)(4.7,-1.0)(4.94,-0.32)(4.72,-2.0)
\pspolygon[linewidth=0.0020,linecolor=lightgray,fillstyle=solid,fillcolor=lightgray](4.72,-2.0)(2.62,-1.18)(2.64,0.08)(3.7,1.0)(4.72,0.6)(4.72,-0.76)
\psbezier[linewidth=0.0020,linecolor=lightgray,fillstyle=solid,fillcolor=lightgray](0.52,0.6)(0.38,0.02)(0.4775044,-0.31918114)(0.48,-0.56)(0.4824956,-0.80081886)(0.42,-1.32)(0.54,-2.0)
\pspolygon[linewidth=0.0020,linecolor=lightgray,fillstyle=solid,fillcolor=lightgray](0.52,0.6)(1.52,1.0)(2.6,0.06)(2.62,-1.16)(0.54,-2.0)(0.52,-0.84)
\psdots[dotsize=0.24](2.62,0.14)
\psline[linewidth=0.076cm](1.92,1.58)(2.6,0.1)
\psline[linewidth=0.076cm](1.54,1.02)(2.6,0.06)
\psline[linewidth=0.076cm](3.28,1.58)(2.6,0.06)
\psline[linewidth=0.076cm](3.68,0.98)(2.6,0.04)
\psline[linewidth=0.076cm](2.62,0.08)(2.62,-1.16)
\psline[linewidth=0.04cm,linecolor=gray](0.52,-2.0)(4.12,-0.6)
\psline[linewidth=0.04cm,linecolor=gray](1.12,-0.6)(4.72,-2.0)
\psline[linewidth=0.04cm,linecolor=gray](0.52,0.6)(1.52,1.0)
\psline[linewidth=0.04cm,linecolor=gray](4.72,0.6)(3.72,1.0)
\psline[linewidth=0.04cm,linecolor=gray](4.32,2.0)(3.32,1.6)
\psline[linewidth=0.04cm,linecolor=gray](0.92,2.0)(1.92,1.6)
\psline[linewidth=0.04cm,linecolor=gray](1.92,1.6)(3.32,1.6)
\psline[linewidth=0.04cm,linecolor=gray](1.52,1.0)(3.72,1.0)
\psline[linewidth=0.04cm](3.72,1.4)(3.72,1.4)
\psline[linewidth=0.04cm,linecolor=gray](1.52,1.0)(1.92,1.6)
\psline[linewidth=0.04cm,linecolor=gray](3.32,1.6)(3.72,1.0)
\psline[linewidth=0.04cm,linecolor=gray](1.52,1.0)(3.32,1.6)
\psline[linewidth=0.04cm,linecolor=gray](1.92,1.58)(2.44,1.42)
\psline[linewidth=0.04cm,linecolor=gray](2.8,1.3)(3.68,1.02)
\psdots[dotsize=0.16,linecolor=gray](2.62,-1.18)
\psdots[dotsize=0.16,linecolor=gray](1.56,1.02)
\psdots[dotsize=0.16,linecolor=gray](3.7,1.0)
\psdots[dotsize=0.16,linecolor=gray](3.3,1.58)
\psdots[dotsize=0.16,linecolor=gray](1.92,1.58)
\psbezier[linewidth=0.0020,fillstyle=solid,fillcolor=lightgray](0.92,0.78)(0.98,0.04)(0.78,0.48)(1.12,-0.6)
\psbezier[linewidth=0.0020,fillstyle=solid,fillcolor=lightgray](4.26,0.78)(4.26,-0.02)(4.42,0.24)(4.12,-0.6)
\end{pspicture} 
\end{minipage}
\begin{minipage}[c]{0.1\textwidth}
{\Huge $\rightarrow$}
\end{minipage}
\begin{minipage}[c]{0.45\textwidth}
\begin{pspicture}(0,-1.821)(4.4,1.82)
\pscircle[linewidth=0.0020,dimen=outer,fillstyle=solid,fillcolor=lightgray](2.6,0.02){1.8}
\psdots[dotsize=0.24](2.6,0.02)
\psline[linewidth=0.04cm](3.72,1.44)(3.72,1.44)
\psline[linewidth=0.05cm,fillcolor=lightgray](2.6,-0.02)(2.6,-1.78)
\psline[linewidth=0.05cm,fillcolor=lightgray](2.6,0.02)(4.32,-0.54)
\psline[linewidth=0.05cm,fillcolor=lightgray](2.6,0.02)(1.54,1.46)
\psline[linewidth=0.03cm,linecolor=gray,fillcolor=gray](4.32,-0.54)(2.6,-1.78)
\psline[linewidth=0.03cm,linecolor=gray,fillcolor=gray](0.88,-0.54)(2.58,-1.78)
\psline[linewidth=0.03cm,linecolor=gray,fillcolor=gray](1.54,1.48)(2.58,-1.76)
\psline[linewidth=0.03cm,linecolor=gray,fillcolor=gray](3.64,1.48)(1.54,1.48)
\psline[linewidth=0.03cm,linecolor=gray,fillcolor=gray](3.64,1.48)(4.32,-0.54)
\psline[linewidth=0.03cm,linecolor=gray,fillcolor=gray](4.3,-0.54)(1.56,1.48)
\psline[linewidth=0.05cm,fillcolor=lightgray](2.6,0.02)(3.66,1.48)
\psline[linewidth=0.05cm,fillcolor=lightgray](2.6,0.02)(0.88,-0.54)
\psline[linewidth=0.03cm,linecolor=gray,fillcolor=gray](1.54,1.48)(0.88,-0.54)
\psline[linewidth=0.03cm,linecolor=gray,fillcolor=gray](3.64,1.48)(0.88,-0.54)
\psline[linewidth=0.03cm,linecolor=gray,fillcolor=gray](3.66,1.48)(2.62,-1.78)
\psline[linewidth=0.03cm,linecolor=gray,fillcolor=gray](4.32,-0.56)(0.88,-0.56)
\psdots[dotsize=0.12,linecolor=gray](0.9,-0.54)
\psdots[dotsize=0.12,linecolor=gray](4.3,-0.54)
\psdots[dotsize=0.12,linecolor=gray](2.6,-1.76)
\psdots[dotsize=0.12,linecolor=gray](1.54,1.46)
\psdots[dotsize=0.12,linecolor=gray](3.64,1.46)
\end{pspicture}  
\end{minipage}
\end{minipage}

\vspace{1.5em}

\noindent (ii) introduce a vertex amplitude $W_v$ to codify the dynamics in the region $B_4$; it is a map from the boundary Hilbert space $\mc{H}_{S^3}$ to $\mbb{C}$; (iii) introduce a state $\Psi_{S^3,q}[s]$ belonging to the boundary Hilbert space $\mc{H}_{S^3}$ to describe the state on $S^3 = \p B_4$.  The role of the state $\Psi_{S^3,q}[s]$ is to codify the dynamics \emph{outside} $B_4$.  As the boundary of the $4$-ball intersects the two-complex giving the complete graph $\Gamma_5$ with five nodes $n_m$ and ten links $l_{mn}$, 
\be
\Gamma_5=
\begin{pspicture}(0,0)(3.5,1.5)
\scalebox{0.7}{
\psline[linewidth=0.03cm](4.32,-0.4303125)(2.6,-1.6703125)
\psline[linewidth=0.03cm](0.88,-0.4303125)(2.58,-1.6703125)
\psline[linewidth=0.03cm](1.54,1.5896875)(2.58,-1.6503125)
\psline[linewidth=0.03cm](3.64,1.5896875)(1.54,1.5896875)
\psline[linewidth=0.03cm](3.64,1.5896875)(4.32,-0.4303125)
\psline[linewidth=0.03cm](4.3,-0.4303125)(1.56,1.5896875)
\psline[linewidth=0.03cm](1.54,1.5896875)(0.88,-0.4303125)
\psline[linewidth=0.03cm](3.64,1.5896875)(0.88,-0.4303125)
\psline[linewidth=0.03cm](3.66,1.5896875)(2.62,-1.6703125)
\psline[linewidth=0.03cm](4.32,-0.4503125)(0.88,-0.4503125)
\psdots[dotsize=0.12](0.9,-0.4303125)
\psdots[dotsize=0.12](4.3,-0.4303125)
\psdots[dotsize=0.12](2.6,-1.6503125)
\psdots[dotsize=0.12](1.54,1.5696875)
\psdots[dotsize=0.12](3.64,1.5696875)
\rput(0.78,-0.60){$n_1$}
\rput(2.55,-1.81){$n_2$}
\rput(4.55,-0.60){$n_3$}
\rput(3.86,1.74){$n_4$}
\rput(1.34,1.74){$n_5$}
\rput(1.58,-1.14){$l_{12}$}
\rput(3.73,-1.14){$l_{23}$}
\rput(4.31,0.64){$l_{34}$}
\rput(2.65,-0.25){$l_{13}$}
\rput(1.85,-0.13){$l_{25}$}
\rput(2.45,0.40){$l_{14}$}
\rput(2.92,0.40){$l_{35}$}
\rput(3.40,-0.13){$l_{24}$}
\rput(0.93,0.64){$l_{15}$}
\rput(2.65,1.82){$l_{45}$}
}
\end{pspicture}
\;,
\ee 
\vspace{3 em}

\noindent the boundary Hilbert space is in fact an $\mc{H}_{\Gamma_5}$ which has the spin networks $|j_{12}\mdots j_{45},i_1\mdots i_5\rangle$ as a basis. Hence, instead of equation (\ref{eq:<4n|H|1n>}), now we have
\be\label{eq:<W|5n>}
W_v(j_{mn},i_n)=
\langle W_v|\!\! 
\begin{pspicture}(0,0)(1.5,.5)
\scalebox{0.3}{
\psline[linewidth=0.03cm](4.32,-0.4303125)(2.6,-1.6703125)
\psline[linewidth=0.03cm](0.88,-0.4303125)(2.58,-1.6703125)
\psline[linewidth=0.03cm](1.54,1.5896875)(2.58,-1.6503125)
\psline[linewidth=0.03cm](3.64,1.5896875)(1.54,1.5896875)
\psline[linewidth=0.03cm](3.64,1.5896875)(4.32,-0.4303125)
\psline[linewidth=0.03cm](4.3,-0.4303125)(1.56,1.5896875)
\psline[linewidth=0.03cm](1.54,1.5896875)(0.88,-0.4303125)
\psline[linewidth=0.03cm](3.64,1.5896875)(0.88,-0.4303125)
\psline[linewidth=0.03cm](3.66,1.5896875)(2.62,-1.6703125)
\psline[linewidth=0.03cm](4.32,-0.4503125)(0.88,-0.4503125)
\psdots[dotsize=0.12](0.9,-0.4303125)
\psdots[dotsize=0.12](4.3,-0.4303125)
\psdots[dotsize=0.12](2.6,-1.6503125)
\psdots[dotsize=0.12](1.54,1.5696875)
\psdots[dotsize=0.12](3.64,1.5696875)
\rput(0.78,-0.60){$i_1$}
\rput(2.55,-1.81){$i_2$}
\rput(4.55,-0.60){$i_3$}
\rput(3.86,1.74){$i_4$}
\rput(1.34,1.74){$i_5$}
\rput(1.58,-1.14){$j_{12}$}
\rput(3.73,-1.14){$j_{23}$}
\rput(4.31,0.64){$j_{34}$}
\rput(2.65,-0.25){$j_{13}$}
\rput(1.85,-0.13){$j_{25}$}
\rput(2.45,0.40){$j_{14}$}
\rput(2.92,0.40){$j_{35}$}
\rput(3.40,-0.13){$j_{24}$}
\rput(0.93,0.64){$j_{15}$}
\rput(2.65,1.82){$j_{45}$}
}
\end{pspicture}
\rangle = \Big(\prod_{m<n} A_f(j_{mn})\Big) A_v(j_{mn},i_n)\;.
\ee
Prescribing the action of the Hamiltonian constraint in this way, i.e. specifying the vertex amplitude, makes it easier to guarantee its crossing symmetry \cite{Reisenberger:1996pu}.\\

The quantities we are interested in here are correlations of geometric operators\footnote{In the following, by $\hat{V}_n$ we mean the volume operator $\hat{V}(\mc{R}_n)$ for a region $\mc{R}_n$ which contains the node $n$. Similarly we denote by $\hat{A}_{mn}$ the area operator $\hat{A}(\mc{S}_{mn})$ for a surface $\mc{S}_{mn}$ which is cut by the link from the node $m$ to the node $n$. These are well-defined operators on $\mc{H}_{\Gamma_5}$.} at the vertex amplitude level, such as area-area correlations on a state $\Psi_{\Gamma_5,q}(j_{mn},i_n)$
\be\label{eq: A A correlation}
\langle\, \hat{A}_{m'n'}\, \hat{A}_{m''n''}\, \rangle_q=\frac{\sum_{j_{mn}} \sum_{i_n} W_v(j_{mn},i_n)\,\hat{A}_{m'n'}\, \hat{A}_{m''n''}\, \Psi_{\Gamma_5,q}(j_{mn},i_n)}{\sum_{j_{mn}} \sum_{i_n} W_v(j_{mn},i_n)\,\Psi_{\Gamma_5,q}(j_{mn},i_n)} 
\ee
or volume-volume correlation $\langle\, \hat{V}_{n'}\, \hat{V}_{n''}\, \rangle_q$. Besides studying two-point correlations, we analyse $n$-point correlations too. Technically, these are correlations of \emph{coloring}, as for instance \cite{Rovelli:1994ge,Ashtekar:1996eg,Ashtekar:1997fb,Ashtekar:1998ak}
\be
\hat{A}_{m'n'} \Psi_{\Gamma_5,q}(j_{mn},i_n)= 8\pi G_N \gamma \sqrt{j_{m'n'}(j_{m'n'}+1)} \Psi_{\Gamma_5,q}(j_{mn},i_n)\;.\label{eq:area eigenvalues}
\ee
The motivation for computing correlations \emph{at the vertex amplitude level} comes from the following key fact: in order to have large scale correlations in a realistic situation where an appropriate boundary semiclassical state is chosen (and a sum over two-complexes is considered), correlations have to be present already at the level of vertex amplitudes. The absence of such correlations would provide an obstruction to the existence of a semiclassical regime. This fact was stressed by Smolin in \cite{Smolin:1996fz}, where the ultra-locality issue was first addressed; see also \cite{Perez:2004hj}, \cite{Nicolai:2005mc,Nicolai:2006id} and \cite{Thiemann:2006cf} for a discussion.

\subsection{Spinfoam vertex amplitudes for quantum gravity} 
Now we analyse some specific models for the dynamics in the Loop Quantum Gravity framework. They are defined giving a vertex amplitude $W_v(j_{mn},i_n)$, i.e. a map from $\mc{H}_{\Gamma_5}$ to $\mbb{C}$. The most elementary ``crossing symmetric'' vertex amplitude is certainly given by $A_v(j_{mn},i_n)=\{15j\}_{SU(2)}$. This vertex amplitude defines a diffeomorphism invariant quantum theory without local degrees of freedom. It corresponds to classical $SU(2)$-$BF$ theory. 

Here we are interested in $4$-dimensional riemannian gravity, which is locally $SO(4)$ invariant. The idea then is to start with a $SO(4)$-invariant weight for the vertex in the two-complex and to take its projection on \emph{a} $SO(3)$ subgroup of $SO(4)$ on the boundary, i.e. on the graph $\Gamma_5$: labelling representations of $SO(4)$ in terms of $SU(2)_L \otimes SU(2)_R$ representations we have
\be
A_v(j_{mn},i_n)=\sum_{j^L_{mn}, i^L_n}\sum_{j^R_{mn}, i^R_n} f^{j_{mn},i_n}_{j^L_{mn}, i^L_n, j^R_{mn}, i^R_n} A_{SO(4)}(j^L_{mn}, i^L_n, j^R_{mn}, i^R_n)\,.
\ee    
A wide class of models can be obtained using the most elementary $SO(4)$-invariant two-complex weight, namely the topological one given by a $SO(4)$-invariant $BF$ theory
\be\label{eq:A_{SO(4)}}
A_{SO(4)}(j_{mn}^L,i_n^L;j_{mn}^R,i_{n}^R)= A_{SU(2)_L}(j_{mn}^L,i_n^L) A_{SU(2)_R}(j_{mn}^R,i_n^R) 
\ee
where $A_{SU(2)}(j_{mn},i_n)$ is given by a $\{15j\}$ symbol. To guarantee crossing symmetry, the branching function 
$f^{j_{mn},i_n}_{j^L_{mn}, i^L_n, j^R_{mn}, i^R_n}$ can be assumed to be factorized over nodes in the following way
\be\label{eq:factorized branching function}
f^{j_{mn},i_n}_{j^L_{mn}, i^L_n, j^R_{mn}, i^R_n}=\Big(\prod_{m<n} c^{j_{mn}}_{j_{mn}^L,j_{mn}^{R}}\Big) \Big(\prod_n f^{i_n}_{i^L_n,i^R_n}\Big)\,.
\ee
Both the Barrett-Crane model \cite{Barrett:1997gw} and the Engle-Pereira-Rovelli model \cite{Engle:2007uq} fall in this class.\\

In the following we use standard diagrammatic methods for $SU(2)$ representation theory\footnote{For an introduction see for instance appendix A of the book \cite{Rovelli:2004tv}, the lectures \cite{Baez:lectures2000} or the book \cite{Cvitanovic:book2007}.}. The $\{15j\}_{SU(2)}$ symbols in formula (\ref{eq:A_{SO(4)}}) can be written introducing a virtual link in representation $k$ to decompose the $4$-valent intertwiners in terms of two $3$-valent ones:

\vspace{-1 em}

\begin{align}
A_{SU(2)}(j_{mn},k_n)=
\begin{pspicture}(0,0)(3,.9)
\scalebox{.5}{ 
\psline[linewidth=0.04cm](3.72,1.42)(3.72,1.42)
\psline[linewidth=0.03cm](4.16,-0.88)(2.92,-1.76)
\psline[linewidth=0.03cm](1.04,-0.88)(2.32,-1.76)
\psline[linewidth=0.03cm](1.28,1.22)(2.32,-1.76)
\psline[linewidth=0.03cm](3.4,1.6)(1.8,1.6)
\psline[linewidth=0.03cm](3.92,1.22)(4.36,-0.28)
\psline[linewidth=0.03cm](4.36,-0.28)(1.8,1.6)
\psline[linewidth=0.03cm](1.28,1.22)(0.84,-0.28)
\psline[linewidth=0.03cm](3.42,1.6)(0.84,-0.28)
\psline[linewidth=0.03cm](3.92,1.22)(2.92,-1.78)
\psline[linewidth=0.03cm](4.16,-0.9)(1.04,-0.9)
\psdots[dotsize=0.12](1.28,1.2)
\psline[linewidth=0.03cm,linestyle=dashed,dash=0.06cm 0.06cm](1.8,1.6)(1.28,1.22)
\psline[linewidth=0.03cm,linestyle=dashed,dash=0.06cm 0.06cm](3.4,1.58)(3.92,1.22)
\psline[linewidth=0.03cm,linestyle=dashed,dash=0.06cm 0.06cm](4.36,-0.28)(4.14,-0.9)
\psline[linewidth=0.03cm,linestyle=dashed,dash=0.06cm 0.06cm](2.32,-1.76)(2.94,-1.76)
\psline[linewidth=0.03cm,linestyle=dashed,dash=0.06cm 0.06cm](1.06,-0.88)(0.82,-0.28)
\psdots[dotsize=0.12](1.8,1.6)
\psdots[dotsize=0.12](3.4,1.6)
\psdots[dotsize=0.12](3.92,1.2)
\psdots[dotsize=0.12](4.36,-0.28)
\psdots[dotsize=0.12](4.14,-0.9)
\psdots[dotsize=0.12](2.94,-1.74)
\psdots[dotsize=0.12](2.3,-1.74)
\psdots[dotsize=0.12](1.06,-0.88)
\psdots[dotsize=0.12](0.84,-0.28)
\rput(0.74,-0.60){$k_1$}
\rput(2.65,-2.00){$k_2$}
\rput(4.55,-0.60){$k_3$}
\rput(3.86,1.65){$k_4$}
\rput(1.34,1.65){$k_5$}
\rput(1.58,-1.44){$j_{12}$}
\rput(3.76,-1.44){$j_{23}$}
\rput(4.41,0.64){$j_{34}$}
\rput(2.65,-0.72){$j_{13}$}
\rput(2.10,-0.17){$j_{25}$}
\rput(2.30,0.50){$j_{14}$}
\rput(2.99,0.50){$j_{35}$}
\rput(3.14,-0.17){$j_{24}$}
\rput(0.8,0.64){$j_{15}$}
\rput(2.65,1.82){$j_{45}$}
}
\end{pspicture}\;.\label{eq:15j splitting}\\[1em]
{}\nonumber
\end{align}
Moreover, we take the following normalization for $3$-valent intertwiners:
\begin{align}
\begin{pspicture}(0,-.1)(1.8,0.65)
\scalebox{.7} 
{
\psbezier[linewidth=0.04](0.08,-0.02921875)(0.48,-0.82921875)(1.68,-0.82921875)(2.08,-0.02921875)
\psbezier[linewidth=0.04](0.08,-0.02921875)(0.68,0.7707813)(1.48,0.7707813)(2.08,-0.02921875)
\psline[linewidth=0.04](0.08,-0.02921875)(2.08,-0.02921875)
\psdots[dotsize=0.16](0.08,-0.02921875)
\psdots[dotsize=0.16](2.08,-0.02921875)
\rput(0.7,0.7){$j_1$}
\rput(1.1,0.2){$j_2$}
\rput(0.7,-0.3){$j_3$}
}
\end{pspicture} =\;\; \left\{
\begin{array}{ll}
1&\textrm{if $j_1,j_2,j_3$ are admissible,}\\
0&\textrm{otherwise.} 
\end{array}
\right.\\[.5em]
{}\nonumber
\end{align}


\subsubsection{The Barrett-Crane model}\label{sec:The Barrett-Crane model}
We restrict attention to the Barrett-Crane model. This model is defined taking a branching function of the form (\ref{eq:factorized branching function}) and defined in the following way: on links we have
\be\label{eq:c branching for BC}
c^{l_{mn}}_{j_{mn}^L,j_{mn}^{R}} = (-1)^{l_{mn}}\,\de_{l_{mn}, (j_{mn}^L+j_{mn}^R)}\; \de_{j_{mn}^L,j_{mn}^R} 
\ee
where we have used the name $l_{mn}$ for the representations on the links of the boundary spin-network. On nodes (using the decomposition introduced in equation (\ref{eq:15j splitting})) we have
\begin{equation}\label{eq:f branching for BC}
f^{k_n}_{k^L_n,k^R_n}=\left\{
\begin{array}{ll}
(-1)^{2 k^L_n} (2 k^L_n+1)\, \de_{k_n^L,k_n^R}
& \textrm{for $k_n$ admissible,}\\[.5 em]
0&\textrm{otherwise.} 
\end{array}
\right.
\end{equation}
Due to the $\de_{j_{mn}^L,j_{mn}^R}$ and the $\de_{k_n^L,k_n^R}$, in this model only \emph{simple} $SO(4)$ representations (on links, as well as on nodes) contribute to the vertex amplitude\footnote{For the relation with General Relativity see \cite{DePietri:1998mb,Perez:2002vg}; for a recent discussion on the appropriateness of taking simple representations on nodes see \cite{Engle:2007uq,Engle:2007qf,Livine:2007vk,Livine:2007ya,Freidel:2007py,Alexandrov:2007pq} and the discussion section at the end of this paper. }. From (\ref{eq:c branching for BC}) one finds that the model is defined only on a subspace of  $\mc{H}_{\Gamma_5}$, the one with integer representations, i.e. it is well defined only for $SO(3)$ spin-network states. In the following we call such representations $l_{mn}$, instead of $j_{mn}$. The sign factor $(-1)^{l_{mn}}$ will be justified later\footnote{It can be compared to the analogous sign factor appearing in \cite{PonzanoRegge:1968} and in \cite{Ooguri:1991ni} for gravity in three dimensions.}. Due to the form of $f^{k_n}_{k^L_n,k^R_n}$, the model is independent from the specific splitting of $SU(2)_L$ and $SU(2)_R$ intertwiners used in (\ref{eq:15j splitting}). Moreover, as $f^{k_n}_{k^L_n,k^R_n}$ is independent of $k_n$, we have that this model weights all the admissible intertwiners with the same weight.

The spinfoam boundary amplitude kernel defined in equation (\ref{eq:<W|5n>}) can then be written as
\be\label{eq:Wv BC}
W_v(l_{mn},i_n)=\Big(\prod_{m<n}(2 l_{mn}+1)^{N_f}\Big) (-1)^{\sum l_{mn}} A_{\textrm{BC}}(l_{mn}/2)\,,
\ee
with the face amplitude generally taken with exponent $N_f=2$. The function $A_{\textrm{BC}}$ in (\ref{eq:Wv BC}) has been introduced in order to make contact with the nomenclature generally used in the literature. We call it the Barrett-Crane vertex function. It was introduced in \cite{Barrett:1997gw} and is defined by the following formula

\vspace{-1em}

\ben
A_{\textrm{BC}}(j_{mn})=\!\!\sum_{k_1\ldots k_5}\Big(\prod_{i=1}^5 (-1)^{2 k_i}(2 k_i+1)\Big) A_{SU(2)_L}(j_{mn},k_n) A_{SU(2)_R}(j_{mn},k_n)\,,
\een
It can be written in diagrammatic notation as follows\footnote{A closed loop gives the (super-)dimension of the representation space $\phantom{\Big|}\begin{pspicture}(0,0)(0.55,0.25)
\scalebox{0.5} 
{
\pscircle[linewidth=0.04,linestyle=dashed,dash=0.06cm 0.06cm,dimen=outer](0.5,0.1){0.5}
}
\end{pspicture}^j=(-1)^{2j} (2j+1)$, i.e. the dimension or minus the dimension depending on the representation being integer or half-integer.}
\begin{align}
A_{\textrm{BC}}(j_{12}\mdots j_{45})=\!\!\sum_{k_1\ldots k_5} \Big(\prod_{i=1}^5 
\begin{pspicture}(0,0)(1.0,1.0)
\scalebox{0.7} 
{
\pscircle[linewidth=0.04,linestyle=dashed,dash=0.06cm 0.06cm,dimen=outer](0.5,0.1){0.5}
\rput(1.2,0.1){$k_i$}
}
\end{pspicture} 
\Big)\!\!\!\!\! \begin{pspicture}(0,0)(3,1.3)
\scalebox{.7}{ 
\psline[linewidth=0.04cm](3.72,1.42)(3.72,1.42)
\psline[linewidth=0.03cm](4.16,-0.88)(2.92,-1.76)
\psline[linewidth=0.03cm](1.04,-0.88)(2.32,-1.76)
\psline[linewidth=0.03cm](1.28,1.22)(2.32,-1.76)
\psline[linewidth=0.03cm](3.4,1.6)(1.8,1.6)
\psline[linewidth=0.03cm](3.92,1.22)(4.36,-0.28)
\psline[linewidth=0.03cm](4.36,-0.28)(1.8,1.6)
\psline[linewidth=0.03cm](1.28,1.22)(0.84,-0.28)
\psline[linewidth=0.03cm](3.42,1.6)(0.84,-0.28)
\psline[linewidth=0.03cm](3.92,1.22)(2.92,-1.78)
\psline[linewidth=0.03cm](4.16,-0.9)(1.04,-0.9)
\psdots[dotsize=0.12](1.28,1.2)
\psline[linewidth=0.03cm,linestyle=dashed,dash=0.06cm 0.06cm](1.8,1.6)(1.28,1.22)
\psline[linewidth=0.03cm,linestyle=dashed,dash=0.06cm 0.06cm](3.4,1.58)(3.92,1.22)
\psline[linewidth=0.03cm,linestyle=dashed,dash=0.06cm 0.06cm](4.36,-0.28)(4.14,-0.9)
\psline[linewidth=0.03cm,linestyle=dashed,dash=0.06cm 0.06cm](2.32,-1.76)(2.94,-1.76)
\psline[linewidth=0.03cm,linestyle=dashed,dash=0.06cm 0.06cm](1.06,-0.88)(0.82,-0.28)
\psdots[dotsize=0.12](1.8,1.6)
\psdots[dotsize=0.12](3.4,1.6)
\psdots[dotsize=0.12](3.92,1.2)
\psdots[dotsize=0.12](4.36,-0.28)
\psdots[dotsize=0.12](4.14,-0.9)
\psdots[dotsize=0.12](2.94,-1.74)
\psdots[dotsize=0.12](2.3,-1.74)
\psdots[dotsize=0.12](1.06,-0.88)
\psdots[dotsize=0.12](0.84,-0.28)
\rput(0.74,-0.60){$k_1$}
\rput(2.65,-2.00){$k_2$}
\rput(4.55,-0.60){$k_3$}
\rput(3.86,1.65){$k_4$}
\rput(1.34,1.65){$k_5$}
\rput(1.58,-1.44){$j_{12}$}
\rput(3.76,-1.44){$j_{23}$}
\rput(4.41,0.64){$j_{34}$}
\rput(2.65,-0.72){$j_{13}$}
\rput(2.10,-0.17){$j_{25}$}
\rput(2.30,0.50){$j_{14}$}
\rput(2.99,0.50){$j_{35}$}
\rput(3.14,-0.17){$j_{24}$}
\rput(0.8,0.64){$j_{15}$}
\rput(2.65,1.82){$j_{45}$}
}
\end{pspicture} \;\;\;\;\otimes\!\!\!\!\!\! \begin{pspicture}(0,0)(3,1.3)
\scalebox{.7}{ 
\psline[linewidth=0.04cm](3.72,1.42)(3.72,1.42)
\psline[linewidth=0.03cm](4.16,-0.88)(2.92,-1.76)
\psline[linewidth=0.03cm](1.04,-0.88)(2.32,-1.76)
\psline[linewidth=0.03cm](1.28,1.22)(2.32,-1.76)
\psline[linewidth=0.03cm](3.4,1.6)(1.8,1.6)
\psline[linewidth=0.03cm](3.92,1.22)(4.36,-0.28)
\psline[linewidth=0.03cm](4.36,-0.28)(1.8,1.6)
\psline[linewidth=0.03cm](1.28,1.22)(0.84,-0.28)
\psline[linewidth=0.03cm](3.42,1.6)(0.84,-0.28)
\psline[linewidth=0.03cm](3.92,1.22)(2.92,-1.78)
\psline[linewidth=0.03cm](4.16,-0.9)(1.04,-0.9)
\psdots[dotsize=0.12](1.28,1.2)
\psline[linewidth=0.03cm,linestyle=dashed,dash=0.06cm 0.06cm](1.8,1.6)(1.28,1.22)
\psline[linewidth=0.03cm,linestyle=dashed,dash=0.06cm 0.06cm](3.4,1.58)(3.92,1.22)
\psline[linewidth=0.03cm,linestyle=dashed,dash=0.06cm 0.06cm](4.36,-0.28)(4.14,-0.9)
\psline[linewidth=0.03cm,linestyle=dashed,dash=0.06cm 0.06cm](2.32,-1.76)(2.94,-1.76)
\psline[linewidth=0.03cm,linestyle=dashed,dash=0.06cm 0.06cm](1.06,-0.88)(0.82,-0.28)
\psdots[dotsize=0.12](1.8,1.6)
\psdots[dotsize=0.12](3.4,1.6)
\psdots[dotsize=0.12](3.92,1.2)
\psdots[dotsize=0.12](4.36,-0.28)
\psdots[dotsize=0.12](4.14,-0.9)
\psdots[dotsize=0.12](2.94,-1.74)
\psdots[dotsize=0.12](2.3,-1.74)
\psdots[dotsize=0.12](1.06,-0.88)
\psdots[dotsize=0.12](0.84,-0.28)
\rput(0.74,-0.60){$k_1$}
\rput(2.65,-2.00){$k_2$}
\rput(4.55,-0.60){$k_3$}
\rput(3.86,1.65){$k_4$}
\rput(1.34,1.65){$k_5$}
\rput(1.58,-1.44){$j_{12}$}
\rput(3.76,-1.44){$j_{23}$}
\rput(4.41,0.64){$j_{34}$}
\rput(2.65,-0.72){$j_{13}$}
\rput(2.10,-0.17){$j_{25}$}
\rput(2.30,0.50){$j_{14}$}
\rput(2.99,0.50){$j_{35}$}
\rput(3.14,-0.17){$j_{24}$}
\rput(0.8,0.64){$j_{15}$}
\rput(2.65,1.82){$j_{45}$}
}
\end{pspicture}\quad{}\,.\label{eq:A_BC}\\[1 em]
{}\nonumber
\end{align}
The occurrence of $l_{mn}/2$ as argument of $A_{\textrm{BC}}$ in (\ref{eq:Wv BC}) can be understood as a consequence of the fact that the generator of $SO(3)$ spatial rotations $L$ can be written in terms of the generators $J_L$ and $J_R$ of $SO(4)\approx SU(2)_L\otimes SU(2)_R$ as $L=J_L+J_R$. While at this level the specific form (\ref{eq:Wv BC}) is simply postulated, at the end of this paper we will give a check of it in a semiclassical regime.

\subsection{The boundary state}\label{sec:The boundary state}
Now we shift attention to the boundary state. As the Barrett-Crane model gives dynamics only to the integer spin subspace of $\mc{H}_{\Gamma_5}$, we restrict to it.  A spinnetwork state $|\Gamma_5, l_{mn}, i_n\rangle \in \mc{H}_{\Gamma_5}$ describes a quantum geometry on a manifold $S^3$. Such quantum geometry consists of $5$ chunks of space (one for each node of the graph $\Gamma_5$), each meeting the other four chunks so that they identify in the whole $10$ patches as prescribed by the connectivity the graph $\Gamma_5$. This picture comes from the fact that the state $|\Gamma_5, l_{mn}, i_n\rangle \in \mc{H}_{\Gamma_5}$ is simultaneously an eigenstate of the volume operator of a region containing a node of $\Gamma_5$ and an eigenstate of the area operator of a surface cut by a link of $\Gamma_5$ \cite{Rovelli:1994ge,Ashtekar:1996eg,Ashtekar:1997fb,Ashtekar:1998ak}. 

Here, we are interested in a state $|\Gamma_5, q \rangle$ on $\mc{H}_{\Gamma_5}$ which is peaked both on the intrinsic and on the extrinsic geometry of $S^3$. As such, we are looking for a (mildly) semiclassical state: it has the same connectivity of the state described above, but it is peaked both on area and on its conjugate momentum and similarly for volume. We take the following ansatz:
\be\label{eq:|Gamma_5,q>}
|\Gamma_5, q \rangle=\sum_{i_n}\sum_{l_{mn}} C f(i_n) \Psi_{l_0,\phi_0}(l_{mn})|\Gamma_5, l_{mn}, i_n\rangle
\ee
with $\Psi_{l_0,\phi_0}(l_{mn})$ given by 
\be\label{eq:gaussian x phase}
\Psi_{l_0,\phi_0}(l_{mn})=\exp\Big(- \frac{1}{2} \sum_{m<n}\sum_{p<q} \alpha_{(mn)(pq)}\frac{(l_{mn}-l^{(0)}_{mn}) (l_{pq}-l^{(0)}_{pq})}{\sqrt{\phantom{\hat{\bar{M}}MM\,}}\hspace{-3.3 em} l^{(0)}_{mn}\: l^{(0)}_{pq}} \Big)\; \exp - i \sum_{m<n} \phi^{(0)}_{mn} l_{mn}\,.
\ee
The matrix $\alpha_{(mn)(pq)}$ is assumed to be real and positive definite in order to suppress large fluctuations of $l_{mn}$ from the mean value $l^{(0)}_{mn}$. As the quantity $\phi^{(0)}_{mn}$ appears only in the phase of (\ref{eq:gaussian x phase}) and $l_{mn}$ is an integer, we can assume without loss of generality that it is in the range $0\leq \phi^{(0)}_{mn} \leq 2 \pi$. It plays the role of mean value of the conjugate momentum to the representation $l_{mn}$. To simplify the analysis, we restrict attention to a symmetric situation with $l^{(0)}_{mn}=l_0$ and $\phi^{(0)}_{mn}=\phi_0$. As stated before, we assume $l_0\gg 1$. Hence, due to the gaussian form of the state which is peaked on the value $l_0$ with dispersion $\sqrt{l_0}$, we have that $l_{mn}$ is essentially restricted to be in the range 
\be\label{eq: l_mn range}
\big(1-\frac{1}{\sqrt{l_0}}\big) l_0 \leq l_{mn} \leq \big(1+\frac{1}{\sqrt{l_0}}\big) l_0\;.
\ee
This peakedness property is a kinematical property; if this state is to be considered semiclassical also in the dynamical sense, it depends on the specific spinfoam model for the dynamics chosen. In the following we discuss this issue for the Barrett-Crane model and show that the request of semiclassicality at a dynamical level fixes the angle $\phi_0$ to a specific value. 

It turns out to be useful to introduce the fluctuation $\de l_{mn}$ and the relative fluctuation $k_{mn}$ defined as
\be
l_{mn}=l_0+\de l_{mn}= (1+k_{mn})\,l_0
\ee
with
\be
k_{mn}=\frac{\de l_{mn}}{l_0}=O(\frac{1}{\sqrt{l_0}})\,.
\ee
The idea of taking coefficients with the simple form of a gaussian times a phase for the $l_{mn}$ part of the semiclassical state may appear naive, but it turns out to be an appropriate choice. We expect the form (\ref{eq:gaussian x phase}) to be quite general, at least for \emph{large} $l_0$, and not specific to the elementary graph $\Gamma_5$ at hand here. Moreover it allows a parametrization of the corrections to the leading behaviour in terms of a polynomial multiplying the gaussian. The polynomial is a function of the relative fluctuation $k_{mn}$ with coefficients to be fixed.

As the Barrett-Crane model is degenerate on intertwiner space, i.e. two states with the same representations $l_{mn}$ on links but different intertwiners at nodes evolve in the same way under the dynamics (\ref{eq:<4n|H|1n>}) defined by the model, we do not discuss the specific form of the coefficients $f(i_n)$ and simply assume that $\sum_{i_n} f(i_n)=1$. The coefficient $C$ in (\ref{eq:|Gamma_5,q>}) has the role of normalization constant. The issue of peakedness on volume and appropriateness of the Barrett-Crane model will be discussed in section \ref{sec:discussion}.\\

Due to the peakedness properties of the boundary semiclassical state, the dominant contribution from the Barrett-Crane vertex amplitude $W_v(l_{mn},i_n)$ to the correlation function (\ref{eq: A A correlation}) comes from the range (\ref{eq: l_mn range}). In the following we give a detailed study of this contribution. We shall present it in terms of the function $A_{\textrm{BC}}(j_{mn})$ defined in (\ref{eq:A_BC}) in order to make the comparison with the existing literature easier.

\section{Integral formulae for the Barrett-Crane vertex}\label{sec:integral formulae}
In this section we give an integral formula for the Barrett-Crane vertex function defined in (\ref{eq:A_BC}). We follow the analysis of \cite{Baez:2002rx} and reproduce the derivation here in order to fix the notation. Using the following formula\footnote{In formula (\ref{eq:BCintertwiner}) the diagrammatic notation $\mc{D}^{(j)}_{mm'}(h)=\quad
\begin{pspicture}(0,-.1)(2.4,0)
\scalebox{.7} 
{
\psline[linewidth=0.04cm](0.0,0)(1.0,0)
\psline[linewidth=0.04cm](1.8,0)(2.8,0)
\psellipse[linewidth=0.04,dimen=outer](1.4,0)(0.4,.2)
\rput(1.4,0){$h$}
\rput(-0.3,-.05){$m$}
\rput(3.1,0){$m'$}
\rput(0.6,.2){$j$}
}
\end{pspicture}$ is used for the group element $h$ in representation $j$.}

\vspace{-2.5em}

\begin{align}
\sum_k 
\begin{pspicture}(0,0)(1.0,1.0)
\scalebox{0.7} 
{
\pscircle[linewidth=0.04,linestyle=dashed,dash=0.06cm 0.06cm,dimen=outer](0.5,0.1){0.5}
\rput(1.2,0.1){$k$}
}
\end{pspicture} \;\;
\begin{pspicture}(0,-.1)(1.4,1.3)
\scalebox{.7} 
{
\psline[linewidth=0.04cm](0.38,1.2)(1.38,0.6)
\psline[linewidth=0.04cm](0.38,0.4)(1.38,0.6)
\psline[linewidth=0.04cm](0.38,-0.4)(1.38,-0.6)
\psline[linewidth=0.04cm](0.38,-1.2)(1.38,-0.6)
\psline[linewidth=0.04cm,linestyle=dashed,dash=0.06cm 0.06cm](1.38,0.6)(1.38,-0.6)
\psdots[dotsize=0.12](1.38,0.6)
\psdots[dotsize=0.12](1.38,-0.6)
\rput(0.13,1.2){$m_1$}
\rput(0.13,0.4){$m_2$}
\rput(0.13,-0.4){$m_3$}
\rput(0.13,-1.2){$m_4$}
\rput(0.93,1.15){$j_1$}
\rput(0.93,0.3){$j_2$}
\rput(0.93,-0.3){$j_3$}
\rput(0.93,-1.1){$j_4$}
\rput(1.6,0.0){$k$}
}
\end{pspicture} \otimes\;\;\;
\begin{pspicture}(0,-.1)(1.4,1.22)
\scalebox{.7} 
{
\psline[linewidth=0.04cm](0.06,0.6)(1.06,1.2)
\psline[linewidth=0.04cm](0.06,0.6)(1.06,0.4)
\psline[linewidth=0.04cm](0.06,-0.6)(1.06,-0.4)
\psline[linewidth=0.04cm](0.06,-0.6)(1.06,-1.2)
\psline[linewidth=0.04cm,linestyle=dashed,dash=0.06cm 0.06cm](0.06,0.6)(0.06,-0.6)
\psdots[dotsize=0.12](0.06,0.6)
\psdots[dotsize=0.12](0.06,-0.6)
\rput(1.31,1.2){$m'_1$}
\rput(1.31,0.4){$m'_2$}
\rput(1.31,-0.4){$m'_3$}
\rput(1.31,-1.2){$m'_4$}
\rput(.51,1.15){$j_1$}
\rput(.51,0.3){$j_2$}
\rput(.51,-0.3){$j_3$}
\rput(.51,-1.1){$j_4$}
\rput(-.2,0.0){$k$}
}
\end{pspicture} 
=\; \int d\mu(h)\quad\;
\begin{pspicture}(0,-.1)(2.88,1.3)
\scalebox{.7} 
{
\psline[linewidth=0.04cm](0.0,1.2)(1.0,1.2)
\psline[linewidth=0.04cm](1.8,1.2)(2.8,1.2)
\psellipse[linewidth=0.04,dimen=outer](1.4,1.2)(0.4,0.2)
\psline[linewidth=0.04cm](0.0,0.4)(1.0,0.4)
\psline[linewidth=0.04cm](1.8,0.4)(2.8,0.4)
\psellipse[linewidth=0.04,dimen=outer](1.4,0.4)(0.4,0.2)
\psline[linewidth=0.04cm](0.0,-0.4)(1.0,-0.4)
\psline[linewidth=0.04cm](1.8,-0.4)(2.8,-0.4)
\psellipse[linewidth=0.04,dimen=outer](1.4,-0.4)(0.4,0.2)
\psline[linewidth=0.04cm](0.0,-1.2)(1.0,-1.2)
\psline[linewidth=0.04cm](1.8,-1.2)(2.8,-1.2)
\psellipse[linewidth=0.04,dimen=outer](1.4,-1.2)(0.4,0.2)
\rput(1.4,1.2){$h$}
\rput(1.4,0.4){$h$}
\rput(1.4,-0.4){$h$}
\rput(1.4,-1.2){$h$}
\rput(-0.3,1.2){$m_1$}
\rput(-0.3,0.4){$m_2$}
\rput(-0.3,-0.4){$m_3$}
\rput(-0.3,-1.2){$m_4$}
\rput(3.1,1.2){$m'_1$}
\rput(3.1,0.4){$m'_2$}
\rput(3.1,-0.4){$m'_3$}
\rput(3.1,-1.2){$m'_4$}
\rput(0.6,1.4){$j_1$}
\rput(0.6,0.6){$j_2$}
\rput(0.6,-0.2){$j_3$}
\rput(0.6,-1.0){$j_4$}
}
\end{pspicture}  
\label{eq:BCintertwiner}\\[0 em]
{}\nonumber
\end{align}
at each of the five nodes of equation (\ref{eq:A_BC}), we can express the Barrett-Crane vertex function as an integral over $SU(2)^5$ \cite{Barrett:1998at}
\be\label{eq:A_BC=prod chi}
A_{\textrm{BC}}(j_{12}\mdots j_{45})=(-1)^{\sum 2 j_{mn}}\int_{SU(2)^5}\prod_{1\leq k\leq 5} d\mu(h_k) \prod_{1\leq m<n \leq 5} \chi^{(j_{mn})}(h_m h_n^{-1})\,.
\ee 
(there is a $-1$ factor for each half-integer loop; notice that this factor is cancelled by the analogous sign factor introduced in (\ref{eq:c branching for BC})). Now we rewrite it as an integral over $(S^3)^5$. A group element $h_k \in SU(2)$ can be described using the spherical coordinates $(\psi_k,\theta_k,\phi_k)$ on the group manifold of $SU(2)$, i.e. on $S^3$. Equivalently, $h_k$ can be described giving a unit-vector  $v_k$ in $\mbb{R}^4$:
\be
v_k=\big(\sin \psi_k \sin \theta_k \sin \phi_k\, ,\, \sin \psi_k \sin \theta_k \cos \phi_k \,,\, \sin \psi_k \cos \theta_k \,,\, \cos \psi_k \big),
\ee  
with the domain for $(\psi_k,\theta_k,\phi_k)$ chosen in the following way
\be
0\leq \psi_k \leq \pi \;\;,\;\;0\leq \theta_k \leq \pi\;\;,\;\;0\leq \phi_k \leq 2\pi\,.
\ee
The Haar measure $d\mu(h_k)$ on $SU(2)$ can be given in terms of the Lebesgue measure on $S^3$
\be
d\Omega(\psi_k,\theta_k,\phi_k)=\sin^2\psi_k \sin\theta_k d\psi_k d\theta_k d \phi_k \,.
\ee
As the $S^3$-solid angle is $\int\! d\Omega=2 \pi^2$, the measure can be normalized to $1$ introducing a factor $(2 \pi^2)^{-1}$ in front of it.

Using Weyl representation formula, the character in representation $j$ of $h_m h^{-1}_n$ appearing in (\ref{eq:A_BC=prod chi}) can be written in terms of the angle between the vectors $v_m$ and $v_n$,
\be\label{eq:(v,v)=Phi}
(v_m,v_n)=\cos \Phi_{mn}\,,
\ee
so that
\be
\chi^{(j_{mn})}(h_m h^{-1}_n)=\frac{\sin (2 j_{mn} + 1) \Phi_{mn}}{\sin \Phi_{mn}}\,.
\ee
The integrand in equation (\ref{eq:A_BC=prod chi}) depends only on scalar products $(v_m,v_n)$ of vectors in $\mbb{R}^4$, hence it is $SO(4)$-invariant. Using this invariance, we can choose coordinates on $S^3$ such that $v_1$ points to the ``north pole'', $v_2$ lies on the ``Greenwich meridian'' with latitude $\psi_2$ and $v_3$ lies on the ``$2$dim-Greenwich meridian'' with $2$-latitude $\psi_3,\theta_3$. 
In this way, we have fixed $SO(4)$ invariance. With this choice of coordinates on $S^3$, the five group elements $h_1\mdots h_5$ are described by the following vectors
\begin{align}
v_1=&\big(0,0,0,1\big),\\
v_2=&\big(0,0,\,\sin\psi_2\,,\,\cos\psi_2\big),\nonumber\\
v_3=&\big(0 ,\, \sin \psi_3 \sin \theta_3  \,,\, \sin \psi_3 \cos \theta_3 \,,\, \cos \psi_3 \big),\nonumber\\
v_4=&\big(\sin \psi_4 \sin \theta_4 \sin \phi_4\, ,\, \sin \psi_4 \sin \theta_4 \cos \phi_4 \,,\, \sin \psi_4 \cos \theta_4 \,,\, \cos \psi_4 \big),\nonumber\\
v_5=&\big(\sin \psi_5 \sin \theta_5 \sin \phi_5\, ,\, \sin \psi_5 \sin \theta_5 \cos \phi_5 \,,\, \sin \psi_5 \cos \theta_5 \,,\, \cos \psi_5 \big),\nonumber
\end{align}
or equivalently by the nine angles $u_i$, with $i=1\ldots 9$
\be
u=(\psi_2,\psi_3,\theta_3,\psi_4,\theta_4,\phi_4,\psi_5,\theta_5,\phi_5).
\ee
Formula (\ref{eq:A_BC=prod chi}) can be written as an integral over $(S^3)^5$
\be
A_{\textrm{BC}}(j_{12},\ldots,j_{45})=(-1)^{\sum 2 j_{mn}}\int_{(S^3)^5}\prod_{1\leq k\leq 5} \frac{d\Omega_k}{2\pi^2} \prod_{1\leq m<n \leq 5} \frac{\sin (2 j_{mn} + 1) \Phi_{mn}}{\sin \Phi_{mn}}\,.
\ee
With the coordinates chosen above, the integrand does not depend on the angles $\psi_1, \theta_1, \phi_1, \theta_2, \phi_2$ and $\phi_3$, so we can integrate over them straightforwardly obtaining a $16 \pi^4$ factor:
\begin{align}
A_{\textrm{BC}}(j_{12},\ldots,j_{45})=&(-1)^{\sum 2 j_{mn}}\frac{16 \pi^4}{(2\pi^2)^5}\int_0^\pi\!\!\!\! d\psi_2 \sin^2 \psi_2  \int_0^\pi\!\!\!\! d\psi_3\int_0^\pi\!\!\!\! d\theta_3 \sin^2 \psi_3 \sin \theta_3 \;\; \times\nonumber\\
\times &    \int_0^\pi\!\!\!\! d\psi_4\int_0^\pi\!\!\!\! d\theta_4\int_0^{2\pi}\!\!\!\! d\phi_4 \sin^2 \psi_4 \sin \theta_4        \int_0^\pi\!\!\!\! d\psi_5\int_0^\pi\!\!\!\! d\theta_5\int_0^{2\pi}\!\!\!\! d\phi_5 \sin^2 \psi_5 \sin \theta_5\;\;\times\nonumber\\
\times & \prod_{1\leq m<n \leq 5} \frac{\sin (2 j_{mn} + 1) \Phi_{mn}(\psi_2,\psi_3,\theta_3,\psi_4,\theta_4,\phi_4,\psi_5,\theta_5,\phi_5)}{\sin \Phi_{mn}(\psi_2,\psi_3,\theta_3,\psi_4,\theta_4,\phi_4,\psi_5,\theta_5,\phi_5)}\,.\label{eq:A_BC=int sin /sin}
\end{align}
The ten angles $\Phi_{mn}$ between the vectors $v_n$, $v_{m}$ in $\mbb{R}^4$ can be written explicitly in terms of the nine angles $u_i$:
\begin{align}\label{eq:Phi_mn(u)}
\Phi_{12}=&\, \psi_2\,,\quad \Phi_{13}= \psi_3\,,\quad \Phi_{14}= \psi_4\,,\quad \Phi_{15}= \psi_5\,,\\
\Phi_{23}=& \cos^{-1}\big(\cos\psi_2 \cos\psi_3 + \cos\theta_3 \sin\psi_2 \sin\psi_3 \big)\,,\nonumber\\
\Phi_{24}=& \cos^{-1}\big(\cos\psi_2 \cos\psi_4 + \cos\theta_4 \sin\psi_2 \sin\psi_4 \big)\,,\nonumber\\
\Phi_{25}=& \cos^{-1}\big(\cos\psi_2 \cos\psi_5 + \cos\theta_5 \sin\psi_2 \sin\psi_5 \big)\,,\nonumber\\
\Phi_{34}=& \cos^{-1}\big(\cos\psi_3 \cos\psi_4 + (\cos\theta_3 \cos\theta_4+\cos\phi_4 \sin\theta_3\sin\theta_4)\sin\psi_3 \sin\psi_4  \big)\,,\nonumber\\
\Phi_{35}=& \cos^{-1}\big(\cos\psi_3 \cos\psi_5 + (\cos\theta_3 \cos\theta_5+ \cos\phi_5 \sin\theta_3\sin\theta_5)\sin\psi_3 \sin\psi_5  \big)\,,\nonumber\\
\Phi_{45}=& \cos^{-1}\big(\cos\psi_4 \cos\psi_5 + (\cos\theta_4 \cos\theta_5+ \cos(\phi_5-\phi_4) \sin\theta_4 \sin\theta_5)\sin\psi_4 \sin\psi_5  \big)\,.\nonumber
\end{align}

Calling $D$ the $9$-dimensional angular domain for the angles $u_i$, formula (\ref{eq:A_BC=int sin /sin}) can be written in the following way
\be\label{eq:A_BC = f(u) sin}
A_{\textrm{BC}}(j_{12},\ldots,j_{45})= (-1)^{\sum 2 j_{mn}}\int_D\prod_{i=1\ldots 9} du_i\, f(u)\, \prod_{m<n} \sin \big( (2 j_{mn}+1) \Phi_{mn}(u) \big)
\ee
where $f(u)$ is given by
\be\label{eq:f(u)}
f(u)=\frac{16 \pi^4}{(2\pi^2)^5} \frac{(\sin u_1)^2\, (\sin u_2)^2 \sin u_3 \,\, (\sin u_4)^2 \sin u_5\, (\sin u_7)^2 \sin u_8}{\prod_{1\leq m<n \leq 5} \sin \Phi_{mn}(u)}.
\ee
Moreover, the product $\prod \sin \big( (2 j_{mn} + 1) \Phi_{mn}(u) \big)$ in formula (\ref{eq:A_BC = f(u) sin}) can be written as a sum of $2^{10}$ terms in the following way
\begin{align}\label{eq:prod sin= signs}
\prod_{1\leq m<n \leq 5} \sin \big( (2 j_{mn} + 1) &\Phi_{mn}(u) \big) = \frac{1}{(2\, i)^{10}} \prod_{m<n} \Big(e^{+i (2 j_{mn} + 1) \Phi_{mn}(u)}-e^{-i (2 j_{mn} + 1) \Phi_{mn}(u)}\Big) = \nonumber \\
& = -\frac{1}{2^{10}}\sum_{b=0}^{2^{10}-1}  (-1)^{\sum b_{mn}} \exp \big( i \sum_{m<n} (-1)^{b_{mn}}(2 j_{mn} + 1) \Phi_{mn}(u) \big)\,,
\end{align}
where we have introduced the integer $b=0\mdots 1023$ and the binary digit notation
\be
b\;\;\rightarrow \;\;(b_{12},b_{13},b_{14},b_{15},b_{23},b_{24},b_{25},b_{34},b_{35},b_{45}) ,
\ee 
with $b_{mn}\in\{0,1\}$ so that
\begin{align*}
b=0\quad\;\: \;\;&\rightarrow\;\; (b_{12},b_{13},b_{14},b_{15},b_{23},b_{24},b_{25},b_{34},b_{35},b_{45})=(0, 0, 0, 0, 0, 0, 0, 0, 0, 0)\\
b=1\quad\;\: \;\;&\rightarrow\;\; (b_{12},b_{13},b_{14},b_{15},b_{23},b_{24},b_{25},b_{34},b_{35},b_{45})=(1, 0, 0, 0, 0, 0, 0, 0, 0, 0)\\
b=2\quad\;\: \;\;&\rightarrow\;\; (b_{12},b_{13},b_{14},b_{15},b_{23},b_{24},b_{25},b_{34},b_{35},b_{45})=(0, 1, 0, 0, 0, 0, 0, 0, 0, 0)\\
b=3\quad\;\: \;\;&\rightarrow\;\; (b_{12},b_{13},b_{14},b_{15},b_{23},b_{24},b_{25},b_{34},b_{35},b_{45})=(1, 1, 0, 0, 0, 0, 0, 0, 0, 0)\\
 &\cdots \\
b=1023\;\; &\rightarrow \;\;(b_{12},b_{13},b_{14},b_{15},b_{23},b_{24},b_{25},b_{34},b_{35},b_{45})
=(1, 1, 1, 1, 1, 1, 1, 1, 1, 1)
\end{align*}
Similarly, the $(-1)^{\sum 2 j_{mn}}$ factor can be written as
\ben
(-1)^{\sum 2 j_{mn}}={\textstyle  \exp \big(- i \sum_{m<n} (-1)^{b_{mn}} (2 j_{mn}+1)\, \pi \big)} .
\een

Notice that, while the integrand in (\ref{eq:A_BC = f(u) sin}) is well defined for configurations $u_i$ such that $\Phi_{mn}(u)=0$ for some $m<n$, the function $f(u)$ is singular at such points. These points correspond to degenerate configurations of the five vectors $v_n$, with two or more of them coinciding. It turns out to be useful to introduce a quantity $A_{\textrm{BC}\eps}(j_{mn})$ defined as the integral on a domain $D_\eps$ such that a ball of radius $\eps$  around degenerate configurations has been excised from $D$. The original formula for the vertex can be obtained taking the limit $\eps \to 0$.

Using formula (\ref{eq:A_BC = f(u) sin}) and formula (\ref{eq:prod sin= signs}), the Barrett-Crane vertex function can be written as (the limit of\footnote{Notice that in (\ref{eq:A_{BC}= lim sum}), while the limit of the sum is well defined, the sum of the limits is not. This amounts to take the Cauchy principal value of the addenda given by (\ref{eq:A^b_{BC}}).}) a  sum of $2^{10}$ terms $A^{(b)}_{\textrm{BC}\eps}$:
\be\label{eq:A_{BC}= lim sum}
A_{\textrm{BC}}(j_{12},\ldots,j_{45})=\lim_{\eps\to 0} -\frac{1}{2^{10}}\sum_{b=0}^{2^{10}-1} (-1)^{\sum b_{mn}} A^{(b)}_{\textrm{BC}\eps}(j_{12},\ldots,j_{45})
\ee
with
\be\label{eq:A^b_{BC}}
A^{(b)}_{\textrm{BC}\eps}(j_{12},\ldots,j_{45})=\int_{D_\eps}\prod_{i=1\ldots 9} du_i\,f(u)\;\exp\big( i \sum_{m<n} (-1)^{b_{mn}} (2 j_{mn} + 1) (\Phi_{mn}(u)+\pi)\big)
\ee
with $f(u)$ defined in (\ref{eq:f(u)}) and $\Phi_{mn}(u)$ in (\ref{eq:(v,v)=Phi}),(\ref{eq:Phi_mn(u)}).  Formulae (\ref{eq:A_{BC}= lim sum}) and (\ref{eq:A^b_{BC}}) for the Barrett-Crane vertex function will have a major role in the analysis of the following two sections.

\section{Dominant contribution from the Barrett-Crane vertex}\label{sec:dominant contribution}
Thanks to the form of the boundary state $\Psi_q(l_{mn},i_n)$ discussed in section \ref{sec:The boundary state} we have that the correlation function (\ref{eq: A A correlation}) can be written in the following way
\begin{align*}
\langle\, &\hat{A}_{m'n'}\; \hat{A}_{m''n''}\, \rangle_q \approx \\
&\;\;\approx (8 \pi G_N \gamma)^2\; l_0^2\; \frac{\displaystyle \int\prod d k_{mn}\; W_v((1+k_{mn})l_0)\, (1+k_{m'n'})\,(1+k_{m''n''})\, \Psi_{l_0,\phi_0}((1+k_{mn})l_0)}{\displaystyle \int\prod d k_{mn}\; W_v((1+k_{mn})l_0)\, \Psi_{l_0,\phi_0}((1+k_{mn})l_0)}\;,
\end{align*}
where we have used the fact that, for $l_0\gg 1$, the sum over $l_{mn}$ can be approximated by an integral over $k_{mn}$ and the eigenvalues of the area can be written as $A_{mn}\approx 8 \pi G_N \gamma l_0 (1+ k_{mn}+O(k^2))$.

Using formula (\ref{eq:A_{BC}= lim sum}) for the Barrett-Crane vertex function, we have that the calculation of correlations of areas can be reduced to computing the following quantity
\be\label{eq:quantity <k_mn k_mn>}
\int \prod_{m<n}d k_{mn} P(k_{mn})\,(-1)^{l_0\sum k_{mn}}\, A^{(b)}_{\textrm{BC}\eps}((1+k_{mn})l_0/2) \, \Psi_{l_0,\phi_0}((1+k_{mn})l_0) 
\ee
with $P(k_{mn})$ given by a polynomial in $k_{mn}$. 

In this section we use the integral formula (\ref{eq:A^b_{BC}}) for the Barrett-Crane vertex function to identify the dominant contribution to the quantity (\ref{eq:quantity <k_mn k_mn>}). Substituting the expression (\ref{eq:A^b_{BC}}) and the semiclassical state (\ref{eq:gaussian x phase}) written in terms of the relative fluctuation $k_{mn}$ and of the mean spin $l_0$ in the quantity (\ref{eq:quantity <k_mn k_mn>}), we have
\begin{align}
&\int \prod_{m<n} d k_{mn} P(k_{mn})\,(-1)^{l_0\sum k_{mn}}\, A^{(b)}_{\textrm{BC}\eps}((1+k_{mn})l_0/2) \, \Psi_{l_0,\phi_0}((1+k_{mn})l_0) =\label{eq:exp(Phimn-phi0)^2}\\
& = \int_{D_\eps} \prod_{i=1}^9 d u_i\, f(u)\, e^{i \sum (-1)^{b_{mn}}\Phi_{mn}(u)} e^{i l_0 \sum \Big((-1)^{b_{mn}}\Phi_{mn}(u)-\phi_0\Big)}  \int \prod_{m<n} d k_{mn} P(k_{mn}) \; \times\nonumber\\
&\times \,C(l_0)\, \exp\Big(- \frac{1}{2} l_0 \sum_{m<n}\sum_{p<q} \alpha_{(mn)(pq)} k_{mn} k_{pq} \Big) \exp i l_0 \sum_{m<n} \Big((-1)^{b_{mn}}\Phi_{mn}(u)-\phi_0\Big) k_{mn}\; \nonumber\\
& = \int_{D_\eps} \prod_{i=1}^9 d u_i\, f(u)\, e^{i \sum (-1)^{b_{mn}}\Phi_{mn}(u)} e^{i l_0 \sum \big((-1)^{b_{mn}}\Phi_{mn}(u)-\phi_0\big)}\,\tilde{C}(l_0)\,P(\frac{i}{l_0} \frac{\p}{\p \phi^{(0)}_{mn}})\, \times\nonumber\\
&\times \exp\Big(- \frac{1}{2} l_0 \sum_{m<n}\sum_{p<q} \alpha^{-1}_{(mn)(pq)}  \Big((-1)^{b_{mn}}\Phi_{mn}(u)-\phi^{(0)}_{mn}\Big) \Big((-1)^{b_{pq}}\Phi_{pq}(u)-\phi^{(0)}_{pq}\Big)  \Big)\; \Big|_{\phi^{(0)}_{mn}=\phi_0}\nonumber
\end{align}
where, in the first equality, we have exchanged the integral over $k_{mn}$ with the integral over the angles $u_i$ while, in the second equality, we have performed the integral over $k_{mn}$. The resulting expression has to be studied in the large $l_0$ case. It turns out to be useful to analyse it in the limit $l_0\to\infty$ the first:
\begin{itemize}
\item The limit $l_0 \to \infty$ and the regular configuration. In the limit we have that the gaussian appearing in the last line of (\ref{eq:exp(Phimn-phi0)^2}) becomes a product of delta functions
\be
\prod_{m<n} \de\Big((-1)^{b_{mn}}\Phi_{mn}(u)-\phi_0\Big)\;.\label{eq:product of delta}
\ee
As $0\leq \Phi_{mn}(u)\leq \pi$ and $0\leq \phi_0\leq 2 \pi$, the delta function is non-zero only for $(-1)^{b_{mn}}=+1$ $\forall m<n$. Thus, in the limit $l_0 \to \infty$,  only the term with $b=0$ in the sum (\ref{eq:A_{BC}= lim sum}) can contribute to the correlation function. Restricting to $b=0$, we have to look for the angles $u_i$ which solve the equation
\be\label{eq:Phimn=phi0}
\Phi_{mn}(u)=\phi_0\quad \forall m<n\,.
\ee 
We find that there are only two solutions. We call them $\bar{u}^0$ and $P\bar{u}_0$. The solution $\bar{u}^0$ is given by
\be
\bar{u}^{0}=\big(\bar{\psi}_2,\bar{\psi}_3,\bar{\theta}_3,\bar{\psi}_4,\bar{\theta}_4,\bar{\phi}_4,\bar{\psi}_5,\bar{\theta}_5,\bar{\phi}_5\big)\,,
\ee
with 
\begin{align}
\bar{\psi}_2=&\,\bar{\psi}_3=\bar{\psi}_4=\bar{\psi}_5=\cos^{-1}(-\frac{1}{4})\quad,\quad \bar{\theta}_3=\bar{\theta}_4=\bar{\theta}_5=\cos^{-1}(-\frac{1}{3})\,,\\
\bar{\phi}_4=&\, 2\pi -\cos^{-1}(-\frac{1}{2})\quad,\quad \bar{\phi}_5 = \cos^{-1}(-\frac{1}{2})\,.\nonumber
\end{align}
This solution corresponds to the following configuration of the five vectors $v_n$
\begin{align}
\bar{v}_1=&(0,0,0,1)\quad,\quad \bar{v}_2=(0,0,\frac{\sqrt{15}}{4},-\frac{1}{4})\quad,\quad \bar{v}_3=(0,\sqrt{5/6},-\frac{\sqrt{5/3}}{4},-\frac{1}{4})\\
\bar{v}_4=&(-\frac{\sqrt{5/2}}{2},-\frac{\sqrt{5/6}}{2},-\frac{\sqrt{5/3}}{4},-\frac{1}{4}) \quad,\quad \bar{v}_5=(\frac{\sqrt{5/2}}{2},-\frac{\sqrt{5/6}}{2},-\frac{\sqrt{5/3}}{4},-\frac{1}{4})\,.\nonumber
\end{align}
These vectors have a $4$-dimensional span and the angle between them is 
\be
\bar{\Phi}_{mn}=\cos^{-1} (\bar{v}_m,\bar{v}_n)=\cos^{-1}(-\frac{1}{4}\,)\;.\label{eq:cos-1 -1/4}
\ee 
Hence we have that, in order for the boundary state to have non-zero correlation functions, the angle $\phi_0$ in the semiclassical state (\ref{eq:gaussian x phase}) has to be fixed to the value $\cos^{-1}(-1/4)$, at least in the limit $l_0 \to \infty$.

The other solution of (\ref{eq:Phimn=phi0}), $P\bar{u}^0$, corresponds to a second set of vectors $\bar{v}_n$ with a $4$-dimensional span and can be obtained from the previous one by a reflection with respect to the $3$-plane identified by the vectors $v_1$, $v_2$ and $v_3$. It corresponds to exchanging $v_4$ with $v_5$, or equivalently to exchanging $\phi_4$ with $\phi_5$. \\

\item Large $l_0$ and the dominant contribution. For large (but finite) $l_0$ we have that, to the integral over $u$, only the $u$ such that
\be
|(-1)^{b_{mn}}\Phi_{mn}(u)-\phi_0|\lesssim \frac{1}{\sqrt{l_0}}
\ee
contribute to the area correlations. As the derivative of $\Phi_{mn}(u)$ at $\bar{u}^0$ and at $P\bar{u}^0$ is of order one, we have that only the term $b=0$ can contribute and only the configuration $u$ such that it belongs to a ball $\mc{B}_{\bar{u}^0}$ centered in $\bar{u}^0$ and of radius $1/\sqrt{l_0}$ or to $\mc{B}_{P\bar{u}^0}$ (defined in a similar way).
\end{itemize}

\noindent Taking into account the analysis presented above, we decompose the angular domain $D_\eps$ in the following way
\be
D_\eps = \mc{B}_{\bar{u}^0} \cup \mc{B}_{P \bar{u}^0}\cup \mc{R}_\eps \;.
\ee
Noticing that, as the integrand in (\ref{eq:A^b_{BC}}) is invariant under exchange of $\bar{\phi}_4$ with $\bar{\phi}_5$, we have 
\be
A^{(0)}_{\textrm{BC}\eps} = 2 A^{(0)}_{\textrm{BC}\mc{B}_{\bar{u}^0}} + A^{(0)}_{\textrm{BC}\mc{R}_\eps}
\ee

\vspace{-.8em}

\noindent and

\vspace{-2em}

\ben
A_{\textrm{BC}}(j_{mn}) = -\frac{1}{2^{10}}\Big( 2 A^{(0)}_{\textrm{BC}\mc{B}_{\bar{u}^0}}(j_{mn}) + \lim_{\eps\to 0} A^{(0)}_{\textrm{BC}\mc{R}_\eps}(j_{mn})+\lim_{\eps\to 0}\sum_{b=1}^{2^{10}-1} (-1)^{\sum b_{mn}} A^{(b)}_{\textrm{BC}\eps}(j_{mn})\Big)\;.
\een
In terms of this decomposition, we have the following result: the terms $A^{(b\neq 0)}_{\textrm{BC}\eps}(j_{mn})$ and $A^{(0)}_{\textrm{BC}\mc{R}_\eps}(j_{mn})$ contribute to correlations only in a exponentially suppressed way:
\be\label{eq:A^b neq0 corr}
\int \prod_{m<n}d k_{mn} P(k_{mn})\, A^{(b\neq 0)}_{\textrm{BC}\eps}((1+k_{mn})l_0/2) \, \Psi_{l_0,\phi_0}((1+k_{mn})l_0) = o(l_0^{-N})\quad \forall N>0\;,
\ee
\be\label{eq:A^0_R corr}
\int \prod_{m<n}d k_{mn} P(k_{mn})\, A^{(0)}_{\textrm{BC}\mc{R}_\eps}((1+k_{mn})l_0/2) \, \Psi_{l_0,\phi_0}((1+k_{mn})l_0) = o(l_0^{-N})\quad \forall N>0\;.
\ee
On the other hand, the term $A^{(0)}_{\textrm{BC}\mc{B}_{\bar{u}^0}}(j_{mn})$ contributes to correlations of areas giving a contribution which is suppressed only as a certain power of $l_0$: 
\be
\int \prod_{m<n}d k_{mn} P(k_{mn})\, A^{(0)}_{\textrm{BC}\mc{B}_{\bar{u}^0}}((1+k_{mn})l_0/2) \, \Psi_{l_0,\phi_0}((1+k_{mn})l_0) = O(l_0^{-\bar{n}})
\ee
for some $\bar{n}$ to be determined.

\section{\texorpdfstring{Large $j_0$ asymptotics of the dominant contribution:\\ stationary phase approximation}{Stationary phase approximation}}\label{sec:stationary phase}
As discussed in the previous section, the dominant contribution to area correlation functions comes from the terms $A^{(0)}_{\textrm{BC}\mc{B}_{\bar{u}^0}}(j_{mn})$ and $A^{(0)}_{\textrm{BC}\mc{B}_{P\bar{u}^0}}(j_{mn})$ of the Barrett-Crane vertex function, and the contributions they give are identical. Here we give a detailed analysis of such contribution studying the asymptotics\footnote{For a detailed discussion of the asymptotics of $A^{(0)}_{\textrm{BC}\mc{R}_\eps}$ and of $A^{(b\neq 0)}_{\textrm{BC}\eps}$ see \cite{Baez:2002rx} and \cite{Barrett:2002ur},\cite{Freidel:2002mj}.} of $A^{(0)}_{\textrm{BC}\mc{B}_{\bar{u}^0}}((1+k_{mn}) j_0)$ for large $j_0$, with fixed $k_{mn}$. We recall that, thanks to the form of the boundary semiclassical state, we have 
\be
|k_{mn}|=\Big|\frac{\de j_{mn}}{j_0}\Big|\lesssim \frac{1}{\sqrt{j_0}} .
\ee
In the following we view the relative fluctuation $k_{mn}$ as a \emph{small} parameter, independent from $j_0$.

The large $j_0$ asymptotic behavior of $A^{(0)}_{\textrm{BC}\mc{B}_{\bar{u}^0}}((1+k_{mn}) j_0)$ can be found using the method of the stationary phase \cite{Erdelyi:1954}: the large $\lambda$ behaviour of a function $F(\lambda,\kappa)$ of the form
\ben
F(\lambda,\kappa)=\int_D \prod_{i=1\ldots d} du_i\,f(u)\;e^{\displaystyle\, i \lambda Q_\kappa(u)}
\een
with $f(u)$ smooth and with compact support in $D$ and with $Q_\kappa(u)$ smooth and having a single isolated stationary point $\bar{u}$ in the interior of $D$, is given by the asymptotic expansion
\be\label{eq:erdelyi}
F(\lambda,\kappa)=\Big(\frac{2 \pi}{\lambda}\Big)^{\!\frac{d}{2}}\;e^{\displaystyle\, i \lambda Q_\kappa(\bar{u})}\;\Big(\sum_{n=0}^N a_n(\bar{u}) \lambda^{-n} \,+\, o(\lambda^{-N})\Big)\,.
\ee 
The coefficients $a_n(\bar{u})$ of the asymptotic series are independent from $\lambda$ and are determined in terms of derivatives of $f(u)$ and $Q_\kappa(u)$ at the stationary point $\bar{u}$. As a consequence, they depend on the parameter $\kappa$. In particular the zero order coefficient $a_0$ is given by
\be\label{eq:a0}
a_0(\bar{u})=\frac{\displaystyle f(\bar{u})\;  e^{\pm i \frac{\pi}{4}}}{\sqrt{\big|\det \big(\big.\frac{\p^2 Q_\kappa}{\p u_i\p u_j}\big|_{\bar{u}}\big)\big|}}
\ee
where the sign in $\pm i \frac{\pi}{4}$ is given by the signature of the Hessian $\big.\frac{\p^2 Q_\kappa}{\p u_i\p u_j}\big|_{\bar{u}}$. For a Feynman-diagram algorithm for the coefficients $a_n(\bar{u})$ see for instance section $5$ of \cite{Zelditch:2001}. If the function $f(u)$ does not vanish on the boundary of the domain $D$, a next-to-leading-order contribution appears. It can be determined by integration by parts: if the phase $Q_\kappa(u_{\p D})$ has a stationary point within $\p D$, an extra contribution $B(\lambda,\kappa)=O(\lambda^{-1}\lambda^{-\frac{d-1}{2}})=O(\lambda^{-\frac{d+1}{2}})$ appears on the right hand side of equation (\ref{eq:erdelyi}).

An asymptotic analysis of $A^{(0)}_{\textrm{BC}\mc{B}_{\bar{u}_0}}(j_{mn})$ is possible thanks to formula   
\be
A^{(0)}_{\textrm{BC}\mc{B}_{\bar{u}_0}}((1+k_{mn}) j_0)= \int_{\mc{B}_{\bar{u}_0}}\prod_{i=1\ldots 9} du_i\,\Big( f(u)\,e^{\displaystyle\, i\, \sum \Phi_{mn}(u)}\Big) \;e^{\displaystyle\, i\, 2 j_0 \sum  (1+ k_{mn}) \big(\Phi_{mn}(u)+\pi\big)}\label{eq:A0 stationary phase}
\ee

\subsection{The perturbative action}
To obtain the large $j_0$ asymptotics of $A^{(0)}_{\textrm{BC}\mc{B}_{\bar{u}_0}}((1+k_{mn}) j_0)$, we have to study the stationary points of the phase $Q_{k}(u)$ defined as
\be\label{eq:phase0}
Q_k(u)=\sum_{m<n} (1+ k_{mn}) \big(\Phi_{mn}(u)+\pi\big)\,,
\ee
i.e., fixed the fluctuations $k_{mn}$, we have to look for angles $\bar{u}_i\in \mc{B}_{\bar{u}^0}$ such that
\be\label{eq:stationary phase}
0=\left.\frac{\p Q_k}{\p u_i}\right|_{\bar{u}}=\sum_{m<n} (1+ k_{mn})\left.\frac{\p \Phi_{mn}}{\p u_i}\right|_{\bar{u}}\,.
\ee
The function $Q_k(\bar{u})$ will play a key role in the determination of a perturbative action $S_{l_0}(\de l_{mn})$ as we shall show in next section.

As noticed in \cite{Barrett:1998gs}, equation (\ref{eq:stationary phase}) has the same form of the Schl\"afli differential identity for a single $4$-simplex having faces of area $(1+k_{mn}) A_0$ (for some given reference scale $A_0$) and dihedral angles $\Phi_{mn}$. Here we do not make use of this fact and look for an explicit expressions for $\Phi_{mn}(\bar{u})$ and $Q_k(\bar{u})$ as functions of the parameter $k_{mn}$. 

Our strategy is the following: we start studying the problem for $k_{mn}=0$, i.e. we look for angles $\bar{u}^{(0)}_i$ in $\mc{B}_{\bar{u}_0}$ such that
\be\label{eq:stationary phase 0}
0=\sum_{m<n} \left.\frac{\p \Phi_{mn}}{\p u_i}\right|_{\bar{u}^{(0)}}\,.
\ee
One can check that $\bar{u}^0$ determined in the previous section as a solution of (\ref{eq:Phimn=phi0}) is an isolated stationary point, i.e.  it is a solution of equation (\ref{eq:stationary phase 0}). Moreover it is the only stationary point in $\mc{B}_{\bar{u}^0}$, i.e. in a ball of radius $1/\sqrt{j_0}=O(k)$ around $\bar{u}^0$. Then, we study perturbation theory in $k_{mn}\ll 1$ around the stationary point.

Given a solution $\bar{u}^0$ of (\ref{eq:stationary phase 0}), the stationary configuration $\bar{u}$ of equation (\ref{eq:stationary phase}) can be determined perturbatively in $k_{mn}$. We write a solution of equation (\ref{eq:stationary phase}) as a series
\be
\bar{u}=\bar{u}^{0}+\bar{u}^{(1)}+\frac{1}{2}\bar{u}^{(2)}+\frac{1}{3!}\bar{u}^{(3)}+\ldots
\ee
with $\bar{u}^{(n)}=O(k^n)$. In particular the linear perturbation $\bar{u}^{(1)}$ is first order in $k_{mn}$. Adopting the notation $\frac{\p \Phi_{mn}}{\p u_i}(u)=\Phi_{mn,i}(u)$, we have that $\bar{u}$ has to satisfy
\be
0=\sum_{m<n}(1+k_{mn})\; \Phi_{mn,i}(\bar{u}^{0}+\bar{u}^{(1)}+\frac{1}{2}\bar{u}^{(2)}+\ldots)
\ee
order by order. Hence, expanding up to $O(k^2)$ and recalling that $\bar{u}^0=O(1)$, $\bar{u}^{(1)}=O(k)$, $\bar{u}^{(2)}=O(k^2)$, we find that $\bar{u}^{0}$, $\bar{u}^{(1)}$ and $\bar{u}^{(2)}$ have to satisfy 
\begin{align}
0=&\, \sum_{m<n}\Phi_{mn,i}(\bar{u}^{0})\,,\label{eq:condit u0}\\
0=& \sum_{m<n} k_{mn}\Phi_{mn,i}(\bar{u}^{0})+ \sum_j \Big(\sum_{m<n} \Phi_{mn,ij}(\bar{u}^{0})\Big) \bar{u}^{(1)}_j\,,\label{eq:condit u1}\\
0=&\,\frac{1}{2}\sum_{j\,l} \Big(\sum_{m<n}\Phi_{mn,ijl}(\bar{u}^{0})\Big) \bar{u}^{(1)}_j \bar{u}^{(1)}_l 
  + \sum_j \Big(\sum_{m<n} k_{mn} \Phi_{mn,ij}(\bar{u}^{0})\Big) \bar{u}^{(1)}_j + \nonumber\\
  &\,+\frac{1}{2} \sum_j \Big(\sum_{m<n} \Phi_{mn,ij}(\bar{u}^{0}) \Big) \bar{u}^{(2)}_j\,.\label{eq:condit u2}
\end{align}
We are interested in the value of the phase $Q_k(u)$ evaluated at a stationary point $\bar{u}$. Expanding up to third order we find
\begin{align*}
Q_k&(\bar{u}^{0}+\bar{u}^{(1)}+\frac{1}{2}\bar{u}^{(2)}+\frac{1}{3!}\bar{u}^{(3)}+\ldots)=\\
=& \sum_{m<n}\big(\Phi_{mn}(\bar{u}^{0})+\pi\big)+\Big(\sum_{m<n} k_{mn} \big(\Phi_{mn}(\bar{u}^{0})+\pi\big) +\sum_i \Big(\sum_{m<n} \Phi_{mn,i}(\bar{u}^{0})\Big) \bar{u}^{(1)}_i\Big)+\\
&+ \frac{1}{2} \Bigg( \sum_{i\,j} \Big(\sum_{m<n}\Phi_{mn,ij}(\bar{u}^{0})\Big) \bar{u}^{(1)}_i \bar{u}^{(1)}_j 
      +2 \sum_i \Big(\sum_{m<n} k_{mn}\Phi_{mn,i}(\bar{u}^{0})\Big)\bar{u}^{(1)}_i  +\\
                  & + \sum_i \Big(\sum_{m<n}\Phi_{mn,i}(\bar{u}^{0})\Big) \bar{u}^{(2)}_i \Bigg)+
\frac{1}{3!} \Bigg( \sum_{i\,j\,l} \Big(\sum_{m<n}  \Phi_{mn,ijl}(\bar{u}^{0})\Big) \bar{u}^{(1)}_i \bar{u}^{(1)}_j \bar{u}^{(1)}_l +\\
                   & +3 \sum_{i\,j} \Big(\sum_{m<n} k_{mn} \Phi_{mn,ij}(\bar{u}^{0})\Big) \bar{u}^{(1)}_i \bar{u}^{(1)}_j 
                    +3 \sum_{i\,j} \Big(\sum_{m<n} \Phi_{mn,ij}(\bar{u}^{0})\Big) \bar{u}^{(1)}_i \bar{u}^{(2)}_j +\\  
                    &+3 \sum_i \Big(\sum_{m<n} k_{mn}\Phi_{mn,i}(\bar{u}^{0})\Big) \bar{u}^{(2)}_i 
                    + \sum_i \Big(\sum_{m<n} \Phi_{mn,i}(\bar{u}^{0})\Big) \bar{u}^{(3)}_i \Bigg)
+ O(k^4). 
\end{align*}
Substituting in the above perturbative expression for $Q_\kappa(\bar{u})$ the conditions (\ref{eq:condit u0}-\ref{eq:condit u2}) on $\bar{u}^{0}$, $\bar{u}^{(1)}$, $\bar{u}^{(2)}$ we have
\begin{align}\label{eq: Q0(u0+u1)}
Q_k&(\bar{u}^{0}+\bar{u}^{(1)}+\frac{1}{2}\bar{u}^{(2)}+\frac{1}{3!}\bar{u}^{(3)}+\ldots)=\\
=&\; \sum_{m<n}\big(\Phi_{mn}(\bar{u}^{0})+\pi\big)
+\sum_{m<n} k_{mn} \big(\Phi_{mn}(\bar{u}^{0})+\pi\big)+
\frac{1}{2}\sum_{i\,j} \Big(-\!\sum_{m<n}\Phi_{mn,ij}(\bar{u}^{0})\Big) \bar{u}^{(1)}_i \bar{u}^{(1)}_j+\nonumber\\
&+\frac{1}{3!} \Bigg(\sum_{i\,j\,l} \Big(\sum_{m<n}  \Phi_{mn,ijl}(\bar{u}^{0})\Big) \bar{u}^{(1)}_i \bar{u}^{(1)}_j \bar{u}^{(1)}_l +
3 \sum_{i\,j}  \Big(\sum_{m<n} k_{mn} \Phi_{mn,ij}(\bar{u}^{0})\Big) \bar{u}^{(1)}_i \bar{u}^{(1)}_j 
                     \Bigg)+ O(k^4) \nonumber
\end{align}
Notice that the dependence on $\bar{u}^{(2)}$ appears only starting from the forth order. This is a straightforward consequence of the function $Q_k(u)$ being evaluated at a stationary point. Hence, to obtain the value of $Q_k(\bar{u})$ up to $O(k^3)$ we need to compute only the correction $\bar{u}^{(1)}$ to the stationary configuration $\bar{u}^0$. 

The correction $\bar{u}^{(1)}$ can be determined in terms of $\bar{u}^{0}$ and of $k_{mn}$ using the linear equation (\ref{eq:condit u1}). Defining the matrix $M_{ij}=\sum_{p<q} \Phi_{pq,ij}(\bar{u}^0)$ and the ``vector''  $N_{(mn)i}=\Phi_{mn,i}(\bar{u}^0)$, we have that the previous equation (\ref{eq:condit u1}) is solved by 
\be\label{eq: sol u1}
\bar{u}^{(1)}_i=\sum_{m<n} W_{(mn)i}\,k_{mn}\quad\textrm{with}\quad W_{(mn)i}=\sum_{j}(M^{-1})_{ij}N_{(mn)j}\;.
\ee 
This gives the stationary configuration $\bar{u}$ to first order in $k_{mn}$:
{\footnotesize
\begin{displaymath}
\bar{u}=\left(\!\!\!\!\!
\begin{array}{c}
\cos^{-1}(-\frac{1}{4})\\
\cos^{-1}(-\frac{1}{4})\\
\cos^{-1}(-\frac{1}{3})\\
\cos^{-1}(-\frac{1}{4})\\
\cos^{-1}(-\frac{1}{3})\\
2\pi\!-\!\cos^{-1}\!(-\frac{1}{2})\\
\cos^{-1}(-\frac{1}{4})\\
\cos^{-1}(-\frac{1}{3})\\
\cos^{-1}(-\frac{1}{2})
\end{array}
\!\!\!\!\!\right)+
\left(\!\!\!\!
\begin{array}{c}
\frac{\sqrt{3/5}}{8}(-18 k_{12} + 7 (k_{13} + k_{14} + k_{15} + k_{23} + k_{24} + k_{25}) - 8 (k_{34} - k_{35} - k_{45}))\\
\frac{\sqrt{3/5}}{8}(- 18 k_{13} + 7 (k_{12} + k_{14} + k_{15} + k_{23}  + k_{34} + k_{35}) - 8 (k_{24} - k_{25} - k_{45}))\\
\frac{1}{2\sqrt{2}}(k_{12} + k_{13} - k_{14} - k_{15}  + 2 (k_{24} + k_{25} + k_{34} + k_{35}) - 4 (k_{23} + k_{45}))\\
\frac{\sqrt{3/5}}{8}(- 18 k_{14} + 7 (k_{12} + k_{13} + k_{15} + k_{24} + k_{34} + k_{45}) - 8( k_{23}  - k_{25}  - k_{35}) )\\
\frac{1}{2\sqrt{2}}(k_{12} - k_{13} + k_{14} - k_{15} + 2 (k_{23} + k_{25} + k_{34} + k_{45}) - 4 (k_{24} + k_{35}) )\\
-\frac{\sqrt{3}}{4}(k_{13} + k_{14}  + k_{23} + k_{24} + k_{35} + k_{45} - 2 (k_{15} + k_{25} + k_{34}))\\
\frac{\sqrt{3/5}}{8}(- 18 k_{15} + 7 (k_{12} + k_{13} + k_{14} + k_{25}  + k_{35} + k_{45}) - 8 (k_{23} + k_{24} + k_{34}))\\
\frac{1}{2\sqrt{2}}(k_{12} - k_{13} - k_{14} + k_{15} + 2 (k_{23} + k_{24} + k_{35} + k_{45}) - 4 (k_{25} + k_{34}) )\\
\frac{\sqrt{3}}{4}(k_{13}  + k_{15} + k_{23}  + k_{25} + k_{34}  + k_{45} - 2 (k_{14} + k_{24} + k_{35}))
\end{array}
\!\!\!\!\right)+O(k^2)\,.
\end{displaymath}}

The stationary configuration to first order in $k_{mn}$ is all what we need to compute a Taylor expansion in $k_{mn}$ of the phase $Q_k(\bar{u}(k_{mn}))$ \emph{up to third order}. Substituting (\ref{eq: sol u1}) in (\ref{eq: Q0(u0+u1)}), we find
\begin{align}
Q_k(\bar{u}(k_{mn}))=&\;Q_0+\sum_{m<n}B_{mn} k_{mn}+\frac{1}{2}\sum_{m<n}\sum_{p<q} K_{(mn)(pq)}k_{mn}k_{pq}+  \label{eq:Qk(u)}\\
&+\frac{1}{3!}\sum_{m<n}\sum_{p<q}\sum_{r<s} I_{(mn)(pq)(rs)}k_{mn}k_{pq}k_{rs}+O(k^4)\nonumber
\end{align}
with
\begin{align}
Q_0=& \sum_{m<n}\big(\Phi_{mn}(\bar{u}^0)+\pi\big)= 10\, \big(\cos^{-1}(-1/4)+\pi \big)\,,\label{eq:Q0}\\
B_{mn}=&\, \Phi_{mn}(\bar{u}^0)+\pi= \cos^{-1}(-1/4)+\pi\,,\label{eq:B()}\\[6pt]
K_{(mn)(pq)}=& \sum_{i\,j}\big(-\!\sum_{u<v}\Phi_{uv,ij}(\bar{u}^0)\big)W_{(mn) i} W_{(pq) j}\,,\label{eq:K()()}\\
I_{(mn)(pq)(rs)}=& \sum_{i\,j\,l} \big(\sum_{u<v}\Phi_{uv,ijl}(\bar{u}^0)\big)W_{(mn) i} W_{(pq) j}W_{(rs) l}+3\sum_{i\,j} \Phi_{mn,ij}(\bar{u}^0) W_{(pq) i} W_{(rs) j}\,.\label{eq:I()()()}
\end{align}
The coefficients $Q_0$ and $B_{mn}$ are easy to determine due to the fact that
\be
\phi_0=\Phi_{mn}(\bar{u}^0)=\cos^{-1}(-1/4)\,,\label{eq:phi0}
\ee
as shown in equation (\ref{eq:cos-1 -1/4}). The coefficients $K_{(mn)(pq)}$ and $I_{(mn)(pq)(rs)}$ can be calculated explicitly. For $K_{(mn)(pq)}$ we have
{\footnotesize
\begin{equation}
K_{(mn)(pq)}=\frac{\sqrt{3/5}}{4}\left(\!\!
\begin{array}{cccccccccc}
-9 & 7/2& 7/2&7/2&   7/2&7/2&   7/2&-4&-4&-4\\
  7/2&-9&   7/2&7/2&   7/2&-4&-4&   7/2&7/2&   -4\\
    7/2&   7/2&-9&7/2&   -4&7/2&-4&   7/2&-4&   7/2\\
     7/2&7/2&   7/2&-9&-4&-4&   7/2&-4&7/2&   7/2\\ 
     7/2&7/2&   -4&-4&-9&7/2&   7/2&7/2&   7/2&-4\\ 
     7/2&-4&   7/2&-4&7/2&   -9&7/2&7/2&   -4&7/2\\ 
     7/2&-4&-4&   7/2&7/2&   7/2&-9&-4&   7/2&7/2\\ 
     -4&7/2&   7/2&-4&7/2&   7/2&-4&-9&   7/2&7/2\\
      -4&7/2&-4&   7/2&7/2&-4&   7/2&7/2&-9&   7/2\\
       -4&-4&7/2&   7/2&-4&7/2&   7/2&7/2&   7/2&-9
       \end{array}
\!\!\right)\label{eq:K LQG}
\end{equation}}

\noindent where the following ordering for the fluctuations has been used, 
\ben
\{k_{mn}\}=(k_{12}, k_{13}, k_{14}, k_{15}, k_{23}, k_{24}, k_{25}, k_{34}, k_{35}, k_{45})\,. 
\een
As expected from the symmetries of the boundary state and the fact that two links belonging to the graph $\Gamma_5$ can be either coincident, or touching at a node or being disjoint, we have that giving the following three components of the matrix $K_{(mn)(pq)}$
\be
K_{(12)(12)}=-\frac{9}{4}\sqrt{\frac{3}{5}}\quad,\quad 
K_{(12)(13)}=+\frac{7}{8}\sqrt{\frac{3}{5}} \quad,\quad
K_{(12)(34)}=-\sqrt{\frac{3}{5}}\;.\label{eq:K()() coeff}
\ee
is enough to reconstruct its structure. The matrix $K_{(mn)(pq)}$ has five negative degenerate eigenvalues $\lambda_{-}=-\sqrt{15}$, four positive degenerate eigenvalues $\lambda_{+}=\sqrt{15}/8$, and a zero eigenvalue $\lambda_0$ corresponding to the eigenvector $v_0=1/\sqrt{10}\, (1,1,1,1,1,1,1,1,1,1)$. The presence of a zero eigenvalue was to be expected due to the fact that a symmetric fluctuation $k_{mn}\equiv k_0$ gives simply a small shift of the large $j_0$.

\noindent Similarly, for the coefficients $I_{(mn)(pq)(rs)}$ there are seven classes of elements:
\begin{align}
I_{(12)(12)(12)}=&-\frac{189}{80}\sqrt{\frac{3}{5}} \quad,\quad 
I_{(12)(12)(13)}=+\frac{347}{160}\sqrt{\frac{3}{5}} \quad,\quad
I_{(12)(12)(34)}=-\frac{14}{5}\sqrt{\frac{3}{5}}\;,\qquad\nonumber\\
I_{(12)(23)(13)}=&-\frac{141}{20}\sqrt{\frac{3}{5}} \quad,\quad
I_{(12)(23)(34)}=+\frac{39}{20}\sqrt{\frac{3}{5}}\;,\label{eq:I()()() coeff}\\
I_{(12)(13)(14)}=&-\frac{453}{160}\sqrt{\frac{3}{5}} \quad,\quad
I_{(12)(23)(45)}=-\frac{3}{10}\sqrt{\frac{3}{5}}\;.\nonumber
\end{align}
All the other components of the tensor $I_{(mn)(pq)(rs)}$ can be obtained by symmetry arguments.

\subsection{The perturbative measure}
Now we shift attention to the term $a_0(\bar{u}(k))$ in the asymptotic expansion (\ref{eq:erdelyi}). Its general expression is given in equation (\ref{eq:a0}). While the analysis above deals with the perturbative action, the following is a derivation of the perturbative measure. 

We evaluate the determinant of the Hessian of $Q_k(u)$ at the stationary point $\bar{u}$, perturbatively in $k_{mn}$. Let's call $H_{ij}$ and $H^0_{ij}$ the $9\times 9$ matrices\footnote{Do not mistake $H^0_{ij}$ for $K_{(mn)(pq)}$. The first is the Hessian of $Q_0(u)$ with respect to the nine angle $u_i$ at the stationary configuration $\bar{u}^0$. On the other hand, $K_{(mn)(pq)}$ is the Hessian of $Q_k(\bar{u})$ with respect to the ten variables $k_{mn}$, at $k_{mn}=0$}
\ben
H_{ij}=\left.\frac{\p^2 Q_k}{\p u_i \p u_j}\right|_{\bar{u}^0+\bar{u}^{(1)}+\ldots}\;,\quad H^{(0)}_{ij}=\left.\frac{\p^2 Q_0}{\p u_i \p u_j}\right|_{\bar{u}^0}\,.
\een
We have that $H_{ij}=H^{(0)}_{ij}+H^{(1)}_{ij}+O(k^2)$, with $H^{(1)}_{ij}=O(k)$. The matrices $H^{(0)}_{ij}$ and $H^{(1)}_{ij}$ can be computed explicitly using the results of the previous section and are given by the following expressions
\ben
H^{(0)}_{ij}=\sum_{m<n}\Phi_{mn,ij}(\bar{u}^0) \quad,\quad
H^{(1)}_{ij}=\sum_{m<n}  \Big(\Phi_{mn,ij}(\bar{u}^0)+\sum_l\sum_{p<q}\Phi_{pq,ijl}(\bar{u}^0) W_{(mn)l} \Big)\,k_{mn}\;.
\een
The matrix $H^0_{ij}$ has four negative non-degenerate eigenvalues and five positive non-degenerate ones. They are all of order one. Then we have that the signatures of $H_{ij}$ and of $H^{(0)}_{ij}$ coincide, as their difference is of order $O(k)$. The determinant of $H_{ij}$ can be computed to first order in $k$ using the following formula:
\begin{align*}
\det H  & = \det \big(H_0 \,(\id + H_0^{-1} H_1 + O(k^2))\,\big)\\
        & = (\det H_0) \big( 1 + \Tr(H_0^{-1} H_1)\big)+O(k^2)\,.
\end{align*} 
An explicit calculation gives
\begin{align}
\left|\det \frac{\p^2 Q_k}{\p u_i \p u_j}\right|_{\bar{u}}= \frac{512}{177147} \sqrt{\frac{5}{3}} \,& \big(\,10\,
-\frac{3}{2} k_{12} + \frac{17}{2} (k_{13}+k_{14}+k_{15})+\\
&\;+11(k_{23}+k_{24}+k_{25}+k_{34}+k_{35}+k_{45})\big)+O(k^2)\,.\nonumber
\end{align}
Similarly, the function $f(u)$ defined in (\ref{eq:f(u)}) can be evaluated perturbatively at the stationary point $\bar{u}$ and we find that it is given by
\begin{align}
f(\bar{u})=\frac{128}{405}\frac{\sqrt{2}}{\pi^6}\Big(1-\frac{1}{10}\Big(&\,\frac{21}{4} k_{12}+\frac{1}{4}(k_{13}+k_{14}+k_{15})+\\
&\;-(k_{23}+k_{24}+k_{25}+k_{34}+k_{35}+k_{45})\Big)\Big) +O(k^2)\,.\nonumber
\end{align}
The ratio appearing in equation (\ref{eq:a0}) is given by
\be
\frac{f(\bar{u})}{\sqrt{\big|\det \big(\big.\frac{\p^2 Q}{\p u_i\p u_j}\big|_{\bar{u}}\big)\big|}}=\frac{12\sqrt{2}}{5 \pi^6}\left(\frac{3}{5}\right)^{\frac{3}{4}}\Bigg( 1- \frac{9}{20}\sum_{m<n}k_{mn}\Bigg)+O(k^2)\,.
\ee
Remarkably, to first order, it is symmetric under exchange of the ten $k_{mn}$. This fact was to be expected on general grounds and provides a consistency check of the calculation. 

The last term in (\ref{eq:A0 stationary phase}) which needs to be computed is $\exp i \sum \Phi_{mn}(\bar{u})$. Using the techniques and the notation of the previous section we have
\begin{align*}
&\sum_{m<n} \Phi_{mn}(\bar{u}^0+\bar{u}^{(1)}+\frac{1}{2}\bar{u}^{(2)}+\frac{1}{3!}\bar{u}^{(3)}+\ldots)\;= \sum_{m<n} \Phi_{mn}(\bar{u}^0)+ \frac{1}{2}\sum_{i\,j}\big(\sum_{m<n}\Phi_{mn,ij}(\bar{u}^0)\big)\bar{u}^{(1)}_i\bar{u}^{(1)}_j+\\
&\quad+\frac{1}{3!}\Big(\!-2\sum_{i\,j\,l}\big(\sum_{m<n}\Phi_{mn,ijl}(\bar{u}^0)\big)\bar{u}^{(1)}_i \bar{u}^{(1)}_j \bar{u}^{(1)}_l-6\sum_{i\,j}\big(\sum_{m<n}k_{mn}\Phi_{mn,ij}(\bar{u}^0)\big)\bar{u}^{(1)}_i\bar{u}^{(1)}_j\Big)+O(k^4)
\end{align*}
Comparing with the definitions (\ref{eq:K()()}),(\ref{eq:I()()()}) and (\ref{eq:phi0}), one finds that
\begin{align}
\sum_{m<n} \Phi_{mn}(\bar{u}(k_{mn}))=&\,\sum_{m<n}\phi_0+\frac{1}{2}\sum_{m<n}\sum_{p<q} (-K_{(mn)(pq)})k_{mn}k_{pq}+\label{eq:sum Phi}\\
&+ \frac{1}{3!}\sum_{m<n}\sum_{p<q}\sum_{r<s} (-2 I_{(mn)(pq)(rs)}) k_{mn}k_{pq}k_{rs}+O(k^4)\nonumber
\end{align}

\bigskip

Having computed all the needed ingredients, we conclude this section giving the asymptotics of the quantity $A^{(0)}_{\textrm{BC}\mc{B}_{\bar{u}_0}}((1+k_{mn})j_0)$: we have that  
\begin{align}
A^{(0)}_{\textrm{BC}\mc{B}_{\bar{u}_0}}((1+k_{mn})j_0)=&\,\frac{e^{i\frac{\pi}{4}} 2^7}{\pi^{3/2}}\Big(\frac{3}{5}\Big)^{\frac{7}{4}}\frac{1}{(2 j_0)^{9/2}}\big(1-\frac{9}{20}\sum_{m<n}k_{mn}+O(k^2)+O(j_0^{-1})\big)\times \label{eq:A0asympt summary}\\
&\times \;e^{\displaystyle i \sum \Phi_{mn}(\bar{u}(k)) }\;e^{\displaystyle i 2 j_0 Q_k(\bar{u}(k))} \;\,+\;\, B(j_0,k_{mn})\nonumber
\end{align}
The functions $Q_k(\bar{u}(k))$ and $\sum \Phi_{mn}(\bar{u}(k))$ contributing to the phase are given in equation  (\ref{eq:Qk(u)}) and (\ref{eq:sum Phi}). The leading contribution to the asymptotic expansion is $O(j_0^{-\frac{9}{2}})$. The correction of order $O(k)$ has been computed explicitly. Corrections of order $O(k^2)$ and  $O(j_0^{-1})$ can be computed using the techniques introduced above: they come from the next-to-leading-order expansion of $a_0(\bar{u}(k))$ and from the leading order of $a_1(\bar{u}(k))$. The extra term $B(j_0,k_{mn})$ appearing in (\ref{eq:A0asympt summary}) is due to the fact that the function $f(u)$ given in equation (\ref{eq:f(u)}) does not vanish on $\p \mc{B}_{\bar{u}_0}$. Such term is of order $O(j_0^{-5})$ and contributes to semiclassical correlation functions only in an exponentially suppressed way as discussed in section \ref{sec:dominant contribution}.

\section{Two-and three-area correlations on a semiclassical state}\label{sec:conclusion LQG}
Now that the dominant contribution from the Barrett-Crane vertex function to correlation formula (\ref{eq: A A correlation}) has been identified 
and computed explicitly
, we are ready to analyse correlations on a semiclassical state. The key result of sections \ref{sec:dominant contribution} and \ref{sec:stationary phase} can be stated as follows: the spinfoam vertex amplitude defined in equation (\ref{eq:Wv BC}) has the following asymptotics for large $l_0$ and $\de l_{mn}$ of order $O(\sqrt{l_0})$
\begin{align}
W_v(l_0+\de l_{mn})=&\, \Big(\prod_{m<n}(2 (l_0 + \de l_{mn})+1)^{N_f}\Big) (-1)^{\sum \de l_{mn}} A_{\textrm{BC}}\big((1+\frac{\de l_{mn}}{l_0}) l_0/2\big)\nonumber\\
=&\, \mc{N}\, \mu_{l_0}(\de l_{mn})\,e^{\displaystyle \, i S_{l_0}(\de l_{mn})}\;+ R(l_0+\de l_{mn})\,.\label{eq:Wv BC asymptotics}
\end{align}
The perturbative action $S_{l_0}(\de l_{mn})$ is given by\footnote{We have called perturbative action the function $S_{l_0}(\de l_{mn})$ due to its similarity to a (lattice) field theoretical action \cite{Montvay:1994cy}. Moreover in the following we use some standard techniques from perturbative quantum field theory, such as splitting the action in a free part and an interaction that can be treated perturbatively in the path integral. However the name action is not completely appropriate and can generate some confusion: an appropriate name for the function $S_{l_0}(\de l_{mn})$ is perturbative Hamilton function as it plays the same role the action evaluated on a classical solution plays in WKB semiclassical perturbation theory.} 
\begin{align}
S_{l_0}(\de l_{mn})=& \phi_0 \sum_{m<n} (l_0 + \de l_{mn}+1) +\frac{1}{2}\sum_{m<n}\sum_{p<q} \frac{K_{(mn)(pq)}}{l_0}\big(1-\frac{1}{l_0}\big) \de l_{mn} \de l_{pq}+ \nonumber \\
&+\frac{1}{3!}\sum_{m<n}\sum_{p<q}\sum_{r<s} \frac{I_{(mn)(pq)(rs)}}{l_0^2}\big(1-\frac{2}{l_0}\big) \de l_{mn} \de l_{pq} \de l_{rs}+O(\de l^4/l_0^3)\;,\label{eq:S_l0}
\end{align}
while the perturbative measure $\mu_{l_0}(\de l_{mn})$ is given by
\be
\mu_{l_0}(\de l_{mn})=l_0^{10(N_f-\frac{9}{20})}\,\Big(1+(N_f-\frac{9}{20})\sum_{m<n}\frac{\de l_{mn}}{l_0}+\ldots\Big)\label{eq:mu_l0}
\ee
with the dots standing for terms of order $O\big((\frac{\de l_{mn}}{l_0})^2\big)$ and $O(l_0^{-1})$ which can be determined using the techniques of the previous section. 

The function $R(l_0+\de l_{mn})$ in equation (\ref{eq:Wv BC asymptotics}) contributes only in a exponentially suppressed way to semiclassical correlation functions as discussed in section \ref{sec:dominant contribution} (see in particular equations (\ref{eq:A^b neq0 corr}) and (\ref{eq:A^0_R corr})).

As large $l_0$ area eigenvalues can be written as 
\be
A_{mn}= 8 \pi G_N \gamma \sqrt{(l_0+\de l_{mn})(l_0+\de l_{mn}+1)}= 8 \pi G_N \gamma\, l_{0}\;\Big(1+\frac{\de l_{mn}}{l_0}+\frac{1}{2l_0}+\ldots\Big)\label{eq:area expansion}
\ee
it turns out to be useful to compute correlations of $\de l_{mn}/l_0$ on the state $\Psi_{\Gamma_5,q}(l_{mn})$ before computing area correlations $\langle\, \hat{A}_{mn}\, \hat{A}_{pq}\, \rangle_q$ and $\langle\, \hat{A}_{mn}\, \hat{A}_{pq}\, \hat{A}_{rs}\,\rangle_q$. We have that the leading contribution to the one-spin and two-spin correlation functions are given by
\begin{align}
\langle \frac{\de l_{mn}}{l_0} \rangle_q=&\; 0+O(1/l_0)\,,\label{eq:one-spin}\\
\langle \frac{\de l_{mn}}{l_0} \frac{\de l_{pq}}{l_0}\rangle_q=&\; \frac{1}{l_0}(iK-\alpha)^{-1}_{(mn)(pq)}+O(1/l_0^2)\;.\label{eq:two-spin}
\end{align}
To compute the three-spin correlation function at leading order, an improvement of the boundary semiclassical state is required. We introduce a perturbative parametrization of the correction needed in terms of a polynomial in $\de l_{mn}/l_0$ multiplying the `free' semiclassical state:
\begin{align}
\Psi_{l_0,\phi_0}(l_0+\de l_{mn})=&\,(1+c_1\sum_{m<n} \frac{\de l_{mn}}{l_0}+\ldots)\exp\Big(- \frac{1}{2} \sum_{m<n}\sum_{p<q} \frac{\alpha_{(mn)(pq)}}{l_0}\de l_{mn} \de l_{pq} \Big)\;\times\nonumber\\
&\times \exp - i \phi_0 \sum_{m<n} (l_0+\de l_{mn})\;.
\end{align}
The leading order correction is parametrized by the single constant $c_1$. The three-spin correlation function can be computed perturbatively viewing the cubic term in the second line of (\ref{eq:S_l0}) as an interaction term which perturbs the free quadratic action. Using Wick's theorem we find:
\begin{align}
&\langle \frac{\de l_{mn}}{l_0} \frac{\de l_{pq}}{l_0} \frac{\de l_{rs}}{l_0}\rangle_q=\, \frac{1}{l_0^2} \sum_{'} i I_{(m'n')(p'q')(r's')} (iK-\alpha)^{-1}_{(m'n')(mn)} (iK-\alpha)^{-1}_{(p'q')(pq)} (iK-\alpha)^{-1}_{(r's')(rs)} + \nonumber\\
&\quad+ \frac{1}{l_0^2} \Big(\sum_{'}\frac{i}{3!} I_{(m'n')(p'q')(r's')} (iK-\alpha)^{-1}_{(m'n')(p'q')} (iK-\alpha)^{-1}_{(r's')(rs)} (iK-\alpha)^{-1}_{(mn)(pq)} + \textrm{perm.}\Big)+\nonumber\\
&\quad+ \frac{1}{l_0^2} (c_1+N_f-\frac{9}{20}) \Big(\sum_{m'<n'} (iK-\alpha)^{-1}_{(m'n')(mn)} (iK-\alpha)^{-1}_{(pq)(rs)} + \textrm{perm.}\Big)+ O(1/l_0^3)\;.\label{eq:three-spin}
\end{align} 
By $\sum_{'}$ we mean the sum over the primed indices appearing in the formula, $\sum_{m'<n'}$ for instance. On the second line of (\ref{eq:three-spin}) nine terms appear corresponding to permutations of the indices of the term written explicitly. Similarly, on the third line three terms appear corresponding to permutations of the indices of the written term. Taking into account the fact that the order $O(1/l_0)$ correction to (\ref{eq:one-spin}) is
\begin{align}
\langle \frac{\de l_{mn}}{l_0}\rangle_q=\,& \frac{1}{l_0}\Big( \big(\sum_{'} \frac{i}{3!}I_{(m'n')(p'q')(r's')} (iK-\alpha)^{-1}_{(p'q')(r's')} (iK-\alpha)^{-1}_{(m'n')(mn)} + \textrm{perm.}\big) +\nonumber\\ 
&\qquad+(c_1+N_f-\frac{9}{20}) \sum_{m'<n'} (iK-\alpha)^{-1}_{(m'n')(mn)}\Big)+O(1/l_0^2)\;.\label{eq:one-spin to order 1/l0}
\end{align}
we have that the joint cumulant for three spin fluctuations (i.e. the connected three-point correlation function) is given by the following simpler expression
\begin{align}
&\langle \frac{\de l_{mn}}{l_0} \frac{\de l_{pq}}{l_0} \frac{\de l_{rs}}{l_0}\rangle_q-\Big(\langle \frac{\de l_{mn}}{l_0} \frac{\de l_{pq}}{l_0}\rangle_q \langle \frac{\de l_{rs}}{l_0}\rangle_q+\textrm{perm.}\Big)+2 \langle \frac{\de l_{mn}}{l_0}\rangle_q \langle\frac{\de l_{pq}}{l_0}\rangle_q \langle\frac{\de l_{rs}}{l_0}\rangle_q=\\
&=\frac{1}{l_0^2} \sum_{'} i I_{(m'n')(p'q')(r's')} (iK-\alpha)^{-1}_{(m'n')(mn)} (iK-\alpha)^{-1}_{(p'q')(pq)} (iK-\alpha)^{-1}_{(r's')(rs)}\,+\,O(1/l_0^3)\;.\nonumber
\end{align}
Using the previous results and the expansion (\ref{eq:area expansion}) we have that the area expectation value and the two-area correlation function to leading order are given by 
\begin{align}
&\hspace{-.5em}\langle A_{mn}\rangle_q = 8 \pi G_N \gamma\, l_{0}\,(1+O(1/l_0))\;,\\
&\hspace{-.5em}\langle A_{mn}\, A_{pq}\rangle_q-\langle A_{mn}\rangle_q \langle A_{pq}\rangle_q= (8 \pi G_N \gamma)^2\, l_{0}^2\;\Big(\frac{(iK-\alpha)^{-1}_{(mn)(pq)}}{l_0}+O(1/l_0^2)\Big)\;.
\end{align}
The three-area correlation function  $\langle\, A_{mn}\,A_{pq}\,A_{rs}\,\rangle_q$ can be computed expanding up to order $O(1/l_0^3)$ and $O(\de l_{mn}^2/l_0^4)$ the expression (\ref{eq:area expansion}) for $A_{mn}$
\begin{align}
A_{mn}=& 8 \pi G_N \gamma l_{0}\Big(1+\frac{1}{2 l_0}-\frac{1}{8 l_0^2}+O(1/l_0^3)\,+\,\big(1+\frac{1}{8 l_0^2}+O(1/l_0^3)\big)\frac{\de l_{mn}}{l_0}+\nonumber\\
       & \hspace{5em}+\big(-\frac{1}{8 l_0^2}+O(1/l_0^3)\big)\big(\frac{\de l_{mn}}{l_0}\big)^2+ O\big(\frac{1}{l_0^2}\big(\frac{\de l}{l_0}\big)^3\big)\,\Big)\nonumber
\end{align}
 and using expressions (\ref{eq:three-spin}) and (\ref{eq:one-spin to order 1/l0}). We have that the three-area joint cumulant is simply given by the following expression
\begin{align}
&\langle A_{mn}\, A_{pq}\,A_{rs}\rangle_q-(\langle A_{mn}\, A_{pq}\rangle_q \langle A_{rs}\rangle_q+\textrm{perm.})+2 \langle A_{mn}\rangle_q \langle A_{pq}\rangle_q \langle A_{rs}\rangle_q=\\
&=(8 \pi G_N \gamma)^3\, l_{0}^3\; \Big(\frac{1}{l_0^2} \sum_{'} i I_{(m'n')(p'q')(r's')} (iK-\alpha)^{-1}_{(m'n')(mn)} (iK-\alpha)^{-1}_{(p'q')(pq)} (iK-\alpha)^{-1}_{(r's')(rs)}+\nonumber\\
&\hspace{8em}+O(1/l_0^3)\Big)\;.\nonumber
\end{align}

\section{Correlations in perturbative quantum Regge-calculus}\label{sec:intro regge calculus}
Our aim in this section is to compute correlations of geometric quantities within the framework of perturbative quantum Regge calculus. The motivation behind it is to compare such correlations with the ones computed from Loop Quantum Gravity in the preceding section. For an introduction to Regge calculus \cite{Regge:1961px} and to the many issues in quantizing it, the reader can refer to the following recent reviews \cite{Loll:1998aj,Hamber:2007fk,Williams:2007zako}. Here we give only a brief introduction. The
attitude we shall adopt is that to consider Regge gravity as ordinary General Relativity but with the metric $g_{\mu\nu}(x)$ restricted to belong to the class of piecewise-flat metrics \cite{Friedberg:1984ma,Feinberg:1984he}, instead of belonging for instance to $C^{\infty}$.

Let's consider a four dimensional manifold $\mc{M}$ with a Riemannian piecewise-flat metric $g_{\mu\nu}(x)$ \cite{Cheeger:1983vq}. A structure of this kind is called a \emph{simplicial manifold} and is completely described by the connectivity of a finite number of vertices and by the lengths of edges connecting two of the vertices. Some nomenclature: we call $C$ the simplicial complex\footnote{There is a restriction a simplicial complex $C$ has to satisfy in order to describe the connectivity of a simplicial manifold. The reader can refer to the lectures \cite{Williams:2007zako} for details.} describing the connectivity of the vertices, $C_{0}=\{v_1,\ldots,v_N\}$ the ensemble of its vertices, $C_{1}$ the ensemble of its edges, $C_{2}$ the ensemble of its triangles, $C_{3}$ the ensemble of its tetrahedra, $C_{4}$ the ensemble of its $4$-simplices. Moreover we call: $e_{ij}$ the edge from the vertex $v_i$ to the vertex $v_j$ and $L_{ij}$ its length, $t_{ijh}$ the triangle with vertices $v_i, v_j, v_h$ and $A_{ijh}$ its area, $T_{ijhk}$ a tetrahedron and $V_{ijhk}$ its volume, $\Sigma_{ijhkl}$ a $4$-simplex and $S_{ijhkl}$ its $4$-volume. Moreover, we call $\theta_{(ijh)kl}$ the dihedral angle between the tetrahedra $T_{ijhk}$ and $T_{ijhl}$ which share the triangle $t_{ijh}$. 

As well-known, the Einstein-Hilbert action for a manifold with boundary is given by \cite{Gibbons:1976ue,Hawking:1996ww} 
\begin{equation}
S[g_{\mu\nu}(x)]=\frac{1}{16 \pi G_N} \int_{\mc{M}} d^4 x \sqrt{g} R + \frac{1}{8 \pi G_N} \int_{\p \mc{M}} d^3 x \sqrt{h} K\,.
\end{equation}
For a piecewise-flat metric $g^{\{C,L_{mn}\}}_{\mu\nu}(x)$, we have that $S[g^{\{C,L_{mn}\}}_{\mu\nu}(x)]=S_{\textrm{Regge}}[C,L_{mn}]$ \cite{Friedberg:1984ma,Feinberg:1984he}, with the Regge action given by \cite{Regge:1961px,Hartle:1981cf}
\begin{align*}
S_{\textrm{Regge}}[C,L_{mn}]=&\;\frac{1}{8 \pi G_N} \sum_{\substack{i<j<h\\ t_{ijh}\in \textrm{int} C_2}} A_{ijh}(L_{mn}) \Big(\,2 \pi -  \sum_{\substack{k<l\\ \hspace{-1em} T_{ijhk},T_{ijhl}\in \textrm{int} C_3 \hspace{-3.5em}}} \theta_{(ijh)kl}(L_{mn})\,\Big) +\\
&+\frac{1}{8 \pi G_N} \sum_{\substack{i<j<h\\ t_{ijh}\in \p C_2}} A_{ijh}(L_{mn}) \Big(\, \pi -  \sum_{\substack{k<l\\ \hspace{-1em} T_{ijhk},T_{ijhl}\in \p C_3 \hspace{-3.5em}}} \theta_{(ijh)kl}(L_{mn})\,\Big)\,.
\end{align*}

The quantum theory can be introduced \`a la Feynman \cite{Misner:1957wq} as a sum over piecewise-flat geometries. This involves a sum over both the connectivity and the lengths, with a specific measure and with an appropriate gauge-fixing procedure in order to avoid multiple counting of a given piecewise-flat geometry. An attitude one can provisionally adopt is to freeze the connectivity to a given simplicial complex $\bar{C}$, and integrate only over the lengths of the edges of $\bar{C}$. Within this setting, the problem of the choice of the measure $\mu_{\bar{C}}(L_{mn})$ has been widely discussed in the literature. Following the point of view we have adhered here, such measure can in principle be determined starting from DeWitt measure for $g^{\{\bar{C},L_{mn}\}}_{\mu\nu}(x)$ and computing the Faddeev-Popov determinant for the change of variables to the invariants $L_{mn}$ \cite{Jevicki:1985ta}\cite{Menotti:1996tm,Menotti:1996ke,Menotti:1996rb,Menotti:1997fi}:
\be\label{eq:FP measure}
\int\ms{D}[g^{\{\bar{C},L_{mn}\}}_{\mu\nu}(x)]\;\delta[\chi\big(g^{\{\bar{C},L_{mn}\}}_{\mu\nu}(x)\big)]\; \Delta^{\!\textrm{FP}}_{\chi}[g^{\{\bar{C},L_{mn}\}}_{\mu\nu}(x)]=\int \Big(\prod_{m<n} d L_{mn}\Big)\; \mu^{\textrm{FP}}_{\bar{C}}(L_{mn})\;.
\ee
The measure $\mu^{\textrm{FP}}_{\bar{C}}(L_{mn})$ turns out to be  highly non-trivial to compute\footnote{For an analysis in two dimensions see \cite{Jevicki:1984gm} and \cite{Menotti:1995ih,Menotti:1995yf,Menotti:1995zw,Menotti:1996de}.} and is expected to be given by a product over $4$-simplices of functions depending only on the single $4$-simplex edge-lengths. On the other hand,  taking the point of view that the measure comes from a discretization of a formal measure in the continuum, a class of ultra-local measures $\mu^{\textrm{ul}}_{\bar{C}}(L_{mn})$ have been proposed and studied numerically\footnote{See for instance \cite{Loll:1998aj}, \cite{Hamber:2007fk} and references therein. See also \cite{Hamber:1997ut}.}: they are given by 
\be\label{eq:ultra local measure}
\int \Big(\prod_{m<n} d L_{mn}\Big)\; \mu^{\textrm{ul}}_{\bar{C}}(L_{mn})= \int \Big(\prod_{m<n} L_{mn}^N d L_{mn}\Big) \; \Theta(L_{mn})
\ee
with the exponent $N$ generally taken to be $+1$ or $-1$. The step function $\Theta(L_{mn})$ in the right hand side of (\ref{eq:ultra local measure}) ensures that the edge lengths satisfy the triangle inequalities and their higher dimensional analogs (see for instance \cite{Wheeler:1964rgt} or appendix \ref{app:regge}). 

As always in a functional integral, boundary conditions have to be specified. Here we assume asymptotic flatness. We follow Ro\v cek and Williams \cite{Rocek:1982fr,Rocek:1982tj} and study the theory perturbatively around a classical solution of Einstein-Regge equations. The classical solution we consider is flat space, i.e. a configuration which has zero deficit angle at each triangle\footnote{In $4d$ the deficit angle at a triangle $t_{ijh}$ is defined as $\eps_{ijh}=2 \pi - \sum_{k<l} \theta_{(ijh)kl}$. It can be shown that the densitized curvature for a piecewise-flat metric is related to the deficit angle in the following way: $\displaystyle \sqrt{g(x)} R(x)= \sum_{t_{ijh}\in C_2}\Big(2\eps_{ijh}\,\int_{\sigma\in t_{ijh}}\hspace{-2em} d^2 \sigma\; \delta^{(4)}(x-\bar{x}(\sigma))\,\Big)$, where $\bar{x}(\sigma)$ is the embedding of the triangle $t_{ijh}$ in $\mc{M}$.}. This amounts to find a flat configuration $(\bar{C},L^0_{mn})$, and quantize perturbatively the fluctuations $\de L_{mn}=L_{mn}-L^0_{mn}$ around this background. The fluctuations are assumed to vanish asymptotically. In this setting, the problem of the choice of functional measure is avoided, at least at the lowest order in perturbation theory, as it turns out to be given by
\be
\mu_{\bar{C}}(L^0_{mn}+\de L_{mn})=\mu_{\bar{C}}(L^0_{mn})\;\Big(1+\sum_{m<n}\mu^{(1)}_{(mn)}\frac{\de L_{mn}}{L^0_{mn}}+\sum_{m<n}\sum_{p<q}\mu^{(2)}_{(mn)(pq)}\frac{\de L_{mn}}{L^0_{mn}}\frac{\de L_{pq}}{L^0_{mn}}\ldots\Big)
\ee
with specific coefficients\footnote{Notice that we have assumed the zeroth order coefficient to be different from zero, $\mu^{(0)}=\mu_{\bar{C}}(L^0_{mn})\neq 0$. This input can be understood in the following way: it is needed in order to match with the continuum functional measure determined perturbatively.} $\mu^{(k)}_{(m_1 n_1)\cdots(m_k n_k)}$. Perturbatively, one can study expectation values of products of localized operators, such as the correlation of volumes of two tetrahedra belonging to $\bar{C}_3$:
\ben
\langle\, \de V_{ijhk}\, \de V_{i'j'h'k'} \rangle_0=\frac{\displaystyle \int\prod_{m<n} d \de L_{mn} \;{\textstyle \Big(1+\sum \mu^{(1)}\frac{\de L}{L^0}+\ldots\Big)}\; \de V_{ijhk}\, \de V_{i'j'h'k'}\; e^{\displaystyle i S_{\bar{C}}(L^0_{mn}+\de L_{mn})}}{\displaystyle \int\prod_{m<n} d \de L_{mn} \;{\textstyle \Big(1+\sum \mu^{(1)}\frac{\de L}{L^0}+\ldots\Big)}\; e^{\displaystyle i S_{\bar{C}}(L^0_{mn}+\de L_{mn})}}
\een 
This integral can be performed as follows. Choose a region $B_4$ which is a subset of $\bar{C_4}$ with the topology of a $4$-ball and such that the tetrahedra $T_{ijhk}$ and $T_{i'j'h'k'}$ belong to its boundary. Call $S_3$ its boundary. The action can be written as a sum over the two regions, $S_{\bar{C}}=S_{B}+S_{\bar{C}\textrm{-}B}$. Then the integrals in the correlation formula can be split in the following way: 
\be
\langle\, \de V_{ijhk}\, \de V_{i'j'h'k'} \rangle_{0}=\frac{\displaystyle \int \hspace{-3.5em}\prod_{\hspace{3.5em} m<n\,,\, e_{mn} \in S_3} \hspace{-3.5em} d \de L_{mn} \;\; W(\de L_{mn})\;\de V_{ijhk}\, \de V_{i'j'h'k'}\;\Psi_0(\de L_{mn})}{\displaystyle \int \hspace{-3.5em}\prod_{\hspace{3.5em} m<n\,,\, e_{mn} \in S_3} \hspace{-3.5em} d \de L_{mn} \;\; W(\de L_{mn})\;\Psi_0(\de L_{mn})}
\ee
where we have defined the functions $W(\de L_{mn})$ and $\Psi_0(\de L_{mn})$
\begin{align}
W(\de L_{mn})=&\int \hspace{-1em}\prod_{\substack{m<n\\ e_{mn}\in \textrm{int} B_4}} \hspace{-1em} d \de L_{mn} \;\;{ \Big(1+\sum_{\substack{m<n\\ e_{mn}\in B_4}} \mu_{B_4\,(mn)}^{(1)}\frac{\de L_{mn}}{L^0}+\ldots\Big)}\;\;  e^{\displaystyle i S_{B}(L^0_{mn}+\de L_{mn})}\,,\\
\Psi_0(\de L_{mn})=&\int \hspace{-2em}\prod_{\substack{m<n\\ \hspace{1em} e_{mn}\in \textrm{int} (\bar{C}\textrm{-}B)}} \hspace{-1.5em} d \de L_{mn} \;\;{ \Big(1+\hspace{-2em}\sum_{\substack{m<n\\ \hspace{1em} e_{mn}\in \textrm{int} (\bar{C}\textrm{-}B)}}\hspace{-2em} \mu_{\bar{C}\textrm{-}B\,(mn)}^{(1)}\frac{\de L_{mn}}{L^0}+\ldots\Big)}\;\;  e^{\displaystyle i S_{\bar{C}\textrm{-}B}(L^0_{mn}+\de L_{mn})}
\end{align}
which depend only on the length-fluctuation of the edges belonging to $S_3$. The function $W(\de L_{mn})$ plays the role of transition amplitude kernel. To lowest order in perturbation theory, it is given by the exponential of $i$ times the Hamilton function, i.e. the perturbative Regge action evaluated on a classical solution corresponding to the boundary condition $(S_3, L^0_{mn}+ \de L_{mn})$. To higher order a non-trivial measure multiplying the exponential is expected to appear. On the other hand, the function $\Psi_0(\de L_{mn})$ describes the state of the system. In the present case it codes the asymptotic flatness boundary condition. The name it deserves is `perturbative vacuum state'. It can be evaluated perturbatively, and to lowest order it is expected to be given by a gaussian peaked on the mean value $\langle \de L_{mn}\rangle=0$ times a phase which codes the mean value of the momentum conjugate to the variable $\de L_{mn}$. As such, it is peaked both on the fluctuation of the intrinsic and of the extrinsic curvature of $S_3$.\\

Here we take as region $B_4$ a single $4$-simplex $\Sigma_{12345}$. In this case we have that the vector of edge-length fluctuations has $10$ components $\{L_{mn}\}=(\de L_{12}\mdots \de L_{45})$. Moreover we assume that the background edge-lengths are all equal, $L^0_{mn}=L_0$, i.e. we have as background the regular $4$-simplex. As declared above, our aim is to compute correlations of areas, volumes, angles, etc. in this setting and compare them with the ones computed at the vertex amplitude level in Loop Quantum Gravity in section \ref{sec:conclusion LQG}. Even if not strictly needed, a possibility is to change variables in the functional integral and choose variables closer to the Loop Quantum Gravity ones: instead of edge-length fluctuations, as variables we choose fluctuations of the areas $\de A_{hkl}$ of the $10$ triangles $t_{hkl}$. In this setting, the transition amplitude kernel is easily written: it is simply given by
\be
W(\de A_{123}\mdots \de A_{345})= \mu(A_0+\de A_{hkl})\; e^{\displaystyle i S_{\textrm{Regge}}(A_0+\de A_{hkl})}
\ee
with $S_{\textrm{Regge}}(A_0+\de A_{hkl})$ given by\footnote{In the case of a single $4$-simplex, for each triangle $t_{ijh}$ there are only two tetrahedra sharing it and hence only one dihedral angle $\theta_{(ijh)kl}$, which we have called $\theta_{(ijh)}$.}
\be
S_{\textrm{Regge}}(A_0+\de A_{hkl})=\frac{1}{8 \pi G_N} \sum_{1 \leq h<k<l \leq 5} (A_0+\de A_{hkl}) \Big(\, \pi - \theta_{(hkl)}\big(L_0+\de L_{mn}(\de A_{rst})\big)\,\Big)\;.\label{eq:S(A0+deA)}
\ee
Notice that in the case of a single $4$-simplex the Regge action and its Hamilton function coincide. Given the symmetries of the background, the function $\mu(A_0+\de A_{hkl})$ to lowest order in perturbation theory is given by
\be
\mu(A_0+\de A_{hkl})=1+\frac{\mu_1}{A_0} \sum_{1\leq h<k<l \leq 5}\hspace{-1em} \de A_{hkl}+\ldots\label{eq:mu regge perturbative}
\ee
On the other hand, the boundary state $\Psi_0(\de A_{hkl})$ is much more difficult to compute from first principles due to its dependence on the connectivity of $\bar{C}\textrm{-}B$ and on the asymptotic condition on the outer boundary. However one can parametrize its form: as discussed above it is expected to be given by
\begin{align}
\Psi_0(\de A_{hkl})=&\,(1+\frac{c_1}{A_0} \sum_{1\leq h<k<l\leq 5}\hspace{-1em} \de A_{hkl}+\ldots ) \;\exp \displaystyle -\frac{1}{2} \frac{1}{8 \pi G_N} \sum_{\substack{1\leq r<s<t\leq 5\\ 1\leq u<v<z\leq 5}}\hspace{-1em} \frac{\alpha_{(rst)(uvz)}(A_0)}{A_0} \, \de A_{rst} \de A_{uvz}\,\times\nonumber\\
&\times \, \exp -i \frac{1}{8 \pi G_N} \sum_{1\leq i<j<h\leq 5}\hspace{-1em} \psi_{ijh}(A_0)\, \de A_{ijh}\,\,.
\end{align}
with $\alpha_{(rst)(uvz)}(A_0)$ and $\psi_{ijh}(A_0)$ dimensionless. The quantity $\psi_{ijh}(A_0)$ codes the mean value of the momentum conjugate to $(8 \pi G_N)^{-1}\de A_{ijh}$. In order to describe flat space, it has to be chosen to its classical value which is given by the derivative with respect to $\de A_{ijh}$ of the Hamilton function (\ref{eq:S(A0+deA)}):
\be
\psi_{ijh}(A_0)\equiv 8 \pi G_N \frac{\p S_{\textrm{Regge}}}{\p A_{ijh}}(A_0)= \pi-\theta_{ijh}(L_0)\,.\label{eq:psi(A0)}
\ee 
The function $\psi_{ijh}(A_0)$ is exactly the angle between the normals to the tetrahedra $T_{ijhk}$ and $T_{ijhl}$ which share the triangle $t_{ijh}$ and belong to a regular $4$-simplex, i.e. it is the Regge version of the extrinsic curvature. It can be easily evaluated (see appendix \ref{app:regge}) and its value is $\cos^{-1}(-1/4)$.

\subsection{The perturbative area-Regge-calculus action}
In this subsection we determine the perturbative expansion of the area-Regge-calculus action for a single $4$-simplex (\ref{eq:S(A0+deA)}). 
We start discussing the background: as shown in \cite{Baez:2002rx} using Bang's theorem, a non-degenerate $4$-simplex which has all its ten triangle-areas equal is necessarily regular, i.e. it has all its edge-lengths equal. Then we need to compute the derivatives of $S_{\textrm{Regge}}(A_0+\de A_{hkl})$ with respect to the area fluctuations. Despite the fact that an explicit formula for $L_{mn}=L_{mn}(A_{123}\mdots A_{345})$ is missing, nevertheless the derivative of $L_{mn}$ with respect to $A_{hkl}$ evaluated at $A_0=\frac{\sqrt{3}}{4} L_0^2$ can be easily found using the inverse function theorem: let's introduce the matrices
\be
U_{(hkl)(mn)}(L_0)= \frac{\p A_{hkl}}{\p L_{mn}}(L_0) \quad\textrm{and} \quad
V_{(mn)(hkl)}(A_0)= \frac{\p L_{mn}}{\p A_{hkl}}(A_0)=(U^{-1})_{(mn)(hkl)} \label{eq:U-V}
\ee
so that we have  $\de L_{mn}= \sum_{h<k<l} V_{(mn)(hkl)} \de A_{hkl}$ (see appendix \ref{app:regge} for details). The first, second and third derivatives of the action (\ref{eq:S(A0+deA)}) with respect to area-fluctuations and evaluated at the regular configuration are given by the following expressions
\begin{align}
\frac{\p S_{\textrm{Regge}}}{\p A_{hkl}}(A_0)\,=&\,\,\frac{1}{8 \pi G_N} \big(\pi-\theta_{hkl}(L_0)\big)=\frac{1}{8 \pi G_N} \cos^{-1}(-\frac{1}{4})\,\,,\label{eq:dS/dA}\\[5pt]
\frac{\p^2 S_{\textrm{Regge}}}{\p A_{hkl} \p A_{rst}}(A_0)\,=&\, -\frac{1}{8 \pi G_N} \sum_{i<j} \frac{\p \theta_{hkl}}{\p L_{ij}}(L_0) \,\,V_{(ij)(rst)}(L_0)\,\,,\label{eq:dS/dAdA}\\[5pt]
\frac{\p^3 S_{\textrm{Regge}}}{\p A_{hkl} \p A_{rst} \p A_{uvz}}(A_0)\,=&\,-\frac{1}{8 \pi G_N} \sum_{m<n}\sum_{p<q}\, V_{(mn)(rst)}(L_0)\,\, V_{(pq)(uvz)}(L_0)\,\,\times\nonumber\\
&\hspace{-7em}\times\,\Big( \frac{\p^2 \theta_{hkl}}{\p L_{mn} \p L_{pq}}(L_0) -
\sum_{i<j} \sum_{a<b<c} \frac{\p \theta_{hkl}}{\p L_{ij}}(L_0) \, V_{(ij)(abc)}(L_0) \,\frac{\p^2 A_{abc}}{\p L_{mn}\p L_{pq} }(L_0)\,\label{eq:dS/dAdAdA}
\Big) \,\,.
\end{align}
In equation (\ref{eq:dS/dA}) we have used the fact that, as a special case of Schl\"afli differential identity, we have
\be
\sum_{1 \leq h<k<l \leq 5}\!\!\! A_0  \left.\frac{\p \theta_{(hkl)}}{\p L_{mn}}\right|_{L_{mn}=L_0}=0\;.
\ee
Using the explicit expressions for the first and second derivative of the area and of the dihedral angle with respect to the edge lengths and the matrix $V_{(mn)(hkl)}(A_0)$ given in appendix \ref{app:regge}, we have evaluated expressions (\ref{eq:dS/dAdA}) and (\ref{eq:dS/dAdAdA}). Thanks to the symmetries of the background and to the fact that two triangles in a $4$-simplex can be either coincident, or sharing an edge, or sharing a single vertex, we have that the matrix (\ref{eq:dS/dAdA}) is completely fixed once the following three components have been computed\footnote{
The matrix (\ref{eq:dS/dAdA}) has a zero eigenvalue with area-fluctuation eigenvector $\{\de A_{hkl}\}$ corresponding to a homogeneous expansion (or contraction) $\de A_{hkl}=\de A_0$. Notice that such homogeneous expansion can be absorbed in the background value, $A_0\to A_0+\de A_0$, which is assumed to be large. For an interpretation of this conformal zero mode as a gauge degree of freedom see \cite{Dittrich:2007wm}.  
}
\begin{align}
\frac{\p^2 S_{\textrm{Regge}}}{\p A_{123} \p A_{123}}(A_0) =&\, \frac{-\frac{9}{4}\sqrt{\frac{3}{5}}}{8 \pi G_N \: A_0} \,\, ,\,\,\,\,
\frac{\p^2 S_{\textrm{Regge}}}{\p A_{123} \p A_{124}}(A_0) = \frac{+\frac{7}{8}\sqrt{\frac{3}{5}}}{8 \pi G_N \: A_0} \,\, ,\,\,\label{eq:dS/dAdA coeff}\\
\frac{\p^2 S_{\textrm{Regge}}}{\p A_{123} \p A_{145}}(A_0) =&\, \frac{-\sqrt{\frac{3}{5}}}{8 \pi G_N \: A_0}\,\,.\nonumber
\end{align}
Similarly, in the case of the third derivatives, we have that the following seven classes of components 

\vspace{-2em}

\begin{align}
 \frac{\p^3 S_{\textrm{Regge}}}{\p A_{123} \p A_{123} \p A_{123}}(A_0) =&\, \frac{-\frac{189}{80}\sqrt{\frac{3}{5}}}{8 \pi G_N\;A_0^2}  \quad , \quad
 \frac{\p^3 S_{\textrm{Regge}}}{\p A_{123} \p A_{123} \p A_{124}}(A_0) = \frac{+\frac{347}{160}\sqrt{\frac{3}{5}}}{8 \pi G_N\;A_0^2} \,\, , \,\,\,\, \label{eq:dS/dAdAdA coeff}\\
 \frac{\p^3 S_{\textrm{Regge}}}{\p A_{123} \p A_{123} \p A_{145}}(A_0) =&\, \frac{-\frac{14}{5}\sqrt{\frac{3}{5}}}{8 \pi G_N\;A_0^2}  \quad , \quad
 \frac{\p^3 S_{\textrm{Regge}}}{\p A_{123} \p A_{124} \p A_{125}}(A_0) = \frac{-\frac{141}{20}\sqrt{\frac{3}{5}}}{8 \pi G_N\;A_0^2} \,\, , \,\,\,\, \nonumber \\
 \frac{\p^3 S_{\textrm{Regge}}}{\p A_{123} \p A_{234} \p A_{345}}(A_0) =&\, \frac{+\frac{39}{20}\sqrt{\frac{3}{5}}}{8 \pi G_N\;A_0^2} \quad , \quad
 \frac{\p^3 S_{\textrm{Regge}}}{\p A_{123} \p A_{124} \p A_{134}}(A_0) = \frac{-\frac{453}{160}\sqrt{\frac{3}{5}}}{8 \pi G_N\;A_0^2} \,\, , \,\,\,\, \nonumber\\ 
 \frac{\p^3 S_{\textrm{Regge}}}{\p A_{123} \p A_{124} \p A_{345}}(A_0) =&\, \frac{-\frac{3}{10}\sqrt{\frac{3}{5}}}{8 \pi G_N\;A_0^2} \quad , \nonumber
\end{align}
completely fix all the components of (\ref{eq:dS/dAdAdA}). Introducing the dimensionless coefficients
\begin{align}
K_{(hkl)(rst)}=&\, 8 \pi G_N \: A_0\: \frac{\p^2 S_{\textrm{Regge}}}{\p A_{hkl} \p A_{rst}}(A_0)\,,\label{eq:K()() Regge}\\
I_{(hkl)(rst)(uvz)}=&\, 8 \pi G_N \: A_0^2\: \frac{\p^3 S_{\textrm{Regge}}}{\p A_{hkl} \p A_{rst} \p A_{uvz}}(A_0)\,,\label{eq:I()()() Regge}
\end{align}
we have that the perturbative expansion in fluctuations of area of the single $4$-simplex action is given by
\begin{align}
S_{\textrm{Regge}}(A_0+\de A_{hkl})=& \frac{1}{8 \pi G_N}\Big(\psi_0 \sum_{h<k<l} (A_0+\de A_{hkl}) + \frac{1}{2} \sum_{h<k<l}\sum_{r<s<t}\frac{K_{(hkl)(rst)}}{A_0} \de A_{hkl} \de A_{rst} +\nonumber\\
&\hspace{-1.5em}+ \frac{1}{3!} \sum_{h<k<l}\sum_{r<s<t}\sum_{u<v<z}\frac{I_{(hkl)(rst)(rst)}}{A_0^2} \de A_{hkl} \de A_{rst} \de A_{uvz}+ O(\frac{\de A^4}{A_0^3}) \;\Big)\,.\label{eq:perturbative area-regge-action}
\end{align}

\subsection{Correlations of area fluctuations: comparison with the Loop Quantum Gravity calculation}\label{sec:comparison LQG RC}
Within the framework introduced above, computing correlations of area fluctuations in perturbative Regge-calculus is straightforward. However, before discussing them, we would like to stress the following facts: 
\begin{itemize}
\item[-] the value of the dihedral angle $\psi_{ijh}(A_0)$ defined in (\ref{eq:psi(A0)}),(\ref{eq:dS/dA}) coincides with the phase $\phi_0$ defined in (\ref{eq:gaussian x phase}) and determined by equations (\ref{eq:product of delta}) and (\ref{eq:cos-1 -1/4});
\item[-] the three numerical coefficients which characterize the matrix $K_{(hkl)(rst)}$ defined in (\ref{eq:K()() Regge}) in terms of the second derivatives with respect to the area of the Regge action determined (\ref{eq:dS/dAdA}) and evaluated in (\ref{eq:dS/dAdA coeff}) coincide with the three coefficients determined on the Loop Quantum Gravity side in equation (\ref{eq:K()() coeff});
\item[-] similarly the seven coefficients (\ref{eq:I()()() Regge}),(\ref{eq:dS/dAdAdA}),(\ref{eq:dS/dAdAdA coeff}) match with the seven coefficients (\ref{eq:I()()() coeff}).
\end{itemize}
Taking into account these facts, we have that the perturbative area-Regge-action $S_{\textrm{Regge}}(A_0+\de A_{hkl})$ defined in (\ref{eq:perturbative area-regge-action}) and the function $S_{l_0}(\de l_{mn})$ introduced in (\ref{eq:S_l0}) to describe a perturbative regime of Loop Quantum Gravity with the dynamics implemented using the Barrett-Crane model \emph{match}, at least up to third order, provided that we identify the spins and the areas in the following way:
\be
l_0 \equiv \frac{A_0}{8 \pi G_N}\qquad \textrm{and} \qquad \de l_{mn} \equiv \frac{\de A_{hkl}}{8 \pi G_N}\;\;.\label{eq:spin=area}
\ee
On the other hand, within the Loop Quantum Gravity framework, the relation between spins and eigenvalues of the area operator is well-known \cite{Rovelli:1994ge,Ashtekar:1996eg,Ashtekar:1997fb}. Using formula (\ref{eq:area expansion}) we find that, at least to lowest order, the relation between spins and areas coming from Loop Quantum Gravity kinematics and the relation we find here at a dynamical semiclassical level (\ref{eq:spin=area}) are in accord\footnote{Notice that the matching between the kinematical relation (\ref{eq:area expansion}) and the dynamical semiclassical one (\ref{eq:spin=area}) requires the Immirzi parameter $\gamma$ to be equal to one. This is expected to be due to the specific form of the Barrett-Crane model for the dynamics and not to be a general fact.}. 

In order to have that the correlations of areas computed in perturbative area-Regge-calculus match with the ones computed in section \ref{sec:conclusion LQG} from Loop Quantum Gravity, we need much more than a matching of actions: a matching of functional measures is needed. This can be accomplished choosing a specific perturbative measure on the Regge calculus side so that it reproduces the perturbative measure (\ref{eq:mu_l0}). To lowest order, this amounts to the choice $\mu_1=N_f-\frac{9}{20}$ in (\ref{eq:mu regge perturbative}).

\section{Discussion}\label{sec:discussion}
The result presented above establishes a correspondence between the non-perturbative dynamics of Loop Quantum Gravity on a semiclassical state and perturbative-Regge-calculus on its vacuum state. The many assumptions behind this result are discussed in section \ref{sec:large-scale} on the Loop Quantum Gravity side and in section \ref{sec:intro regge calculus} on the Regge calculus side. A fact we would like to stress here is that such correspondence is a non-trivial result, in the sense that it cannot be established on general grounds without working out the details as done above. Some steps in the analysis that a priori can go wrong - but actually turn out to work - are the following:

\begin{itemize}
\item the correspondence has been shown to hold for a specific model for the dynamics in Loop Quantum Gravity - the Barrett-Crane model - but there is no guarantee that it keeps on holding for other models for the dynamics. Obstructions can arise at two levels:
\begin{itemize}
\item[(i)] at the non-perturbative level, due to a different dynamics in the $4$-valent spin-networks sector. For instance, using the Hamiltonian constraint introduced by Thiemann in \cite{Thiemann:1996aw,Thiemann:1996av}\cite{Borissov:1997ji} we have that the matrix elements (\ref{eq:<4n|H|1n>}) turn out to be zero. Therefore, no matching with Regge calculus is to be expected;
\item[(ii)] at the `semiclassical' level, due to a suppression of spin-spin correlations. Before \cite{Rovelli:2005yj,Bianchi:2006uf}, the Barrett-Crane model was suspected to suffer from this pathology. Within the framework discussed in the present paper, this scenario can be stated as follows: there is no superposition of spin-network states with graph $\Gamma_5$ as in (\ref{eq:|Gamma_5,q>}) such that the two-spin correlation function is different from zero and of order $O(1/l_0)$ as in (\ref{eq:two-spin}). In the case of the Barrett-Crane model we have found that this dynamical requirement fixes the phase $\phi_0$ in the boundary semiclassical state (see equation (\ref{eq:product of delta})). At a following stage, we have compared such phase with the dihedral angle appearing in (\ref{eq:psi(A0)}) and found the two to match. On the other hand, assuming the correspondence with Regge calculus, one could start with a superposition of spin-network states such that, at the kinematical level, it describes a state peaked on the intrinsic and the extrinsic $3$-geometry of the boundary of a regular $4$-simplex as done in \cite{Rovelli:2005yj,Bianchi:2006uf}. Then, the calculation of spin-spin correlations comes as a test of semiclassicality at the dynamical level. If these large scale correlations are present for the recently proposed new models for the dynamics \cite{Engle:2007uq,Engle:2007qf,Livine:2007vk,Livine:2007ya,Freidel:2007py,Alexandrov:2007pq} is still an open issue. The same remark is expected to hold also when, instead of using the spinfoam formalism, the dynamics is implemented using canonical methods as for the graph changing Master constraint \cite{Thiemann:2005zg}. 
\end{itemize}
\item in our analysis we have chosen an highly symmetric semiclassical state supported by a graph $\Gamma_5$ which is dual to the boundary of a $4$-simplex. Provided that obstructions (i) and (ii) are avoided so that the dynamics turns out to be effectively described by a perturbative action like the one in equation (\ref{eq:S_l0}), a question one could ask is if the $10\times 10$ matrix $K_{(mn)(pq)}$ and the $10\times 10\times 10$ tensor $I_{(mn)(pq)(rs)}$ in (\ref{eq:S_l0}) are completely fixed by symmetry so that their matching with the ones appearing on the Regge calculus side in equation (\ref{eq:perturbative area-regge-action}) can be predicted a priori. The answer is clearly no as one can respect symmetry and still have $3$ free parameters at the level of $K_{(mn)(pq)}$ and $7$ free parameters at the level of $I_{(mn)(pq)(rs)}$. The reason of the matching comes from the correspondence of the stationary phase equation (\ref{eq:stationary phase}) with the Schl\"afli differential identity for a single $4$-simplex (\ref{eq:Schl id}). Hence we expect on general grounds that the matching keeps on holding at the following orders in the Taylor expansion. What is non-trivial here is the matching of dimensionful parameters, namely the $(8 \pi G_N)^{-1}$ in front of the perturbative action. In fact Schl\"afli identity cannot fix the overall scale. The correct dimensions come from the fact that in (\ref{eq:S_l0}) the dominant terms are of the form $\de l_{mn}^k/l_0^{k-1}$, together with the relation (\ref{eq:area eigenvalues}) between spin and area eigenvalues confirmed in (\ref{eq:spin=area}). On the other hand, the subleading contribution coming from the term (\ref{eq:sum Phi}) gives a correction which amounts to a shift $l_0\to l_0+1$.
\item in our analysis we have computed perturbatively the measure $\mu_{l_0}(\de l_{mn})$ and found that to lowest order it is given by a non-vanishing constant plus a linear term in $\de l_{mn}/l_0$ with a specific coefficient (see equation (\ref{eq:mu_l0})). While on the Regge calculus side there is some uncertainty on the choice of the appropriate measure, the fact that perturbatively it starts with a non-vanishing constant is certainly a requirement if one wishes to make contact with perturbative quantum gravity in the continuum (see for instance \cite{Hamber:1997ut} for a discussion and an example of pathological measure in quantum Regge calculus).  
\end{itemize}
Moreover, using the techniques introduced in sections \ref{sec:dominant contribution} and \ref{sec:stationary phase}, our analysis can be extended to higher orders in perturbation theory. A matter of interest here is the study of higher order corrections to the measure $\mu_{l_0}(\de l_{mn})$. A detailed analysis could clarify the following issues:
\begin{itemize}
\item[-] going at least to quadratic order in $\de l_{mn}$ would allow to check if the perturbative measure coming from Loop Quantum Gravity matches with the expansion of the diffeomorphism-invariant measure (\ref{eq:FP measure}) $\mu^{\textrm{FP}}_{4\textrm{-sim}}(A_0+\de A_{hkl})$ determined on the quantum-Regge-calculus side following \cite{Jevicki:1985ta} and \cite{Menotti:1996tm,Menotti:1996ke,Menotti:1996rb,Menotti:1997fi}, or not. We argue that something new about diffeomorphism invariance in Loop Quantum Gravity can be learned from this comparison; 
\item[-] the perturbative expansion of the coefficients $a_n(\bar{u}(k_{mn})) l_0^{-n}$ appearing in the expression (\ref{eq:erdelyi}) for the asymptotic expansion can be exponentiated giving an `effective action' $S^{\textrm{eff}}_{l_0}(\de l_{mn})$. On the other hand, the result (\ref{eq:spin=area}) provides a dictionary between spins and dimensionful quantities. By dimensional analysis we have that from $a_0$ can come only terms independent of $G_N$, from $a_1$ only terms which scale as $8 \pi G_{N}$ and similarly for higher orders. This suggests that the contribution to the perturbative action coming from the measure $\mu_{l_0}(\de l_{mn})$ can be written in terms of an higher-curvature Regge action \cite{Hamber:1984kx,Hamber:1985gx}, with specific coefficients for the higher-curvature terms fixed by the non-perturbative theory. While suggestive, this conjecture needs a detailed analysis\footnote{A preliminary analysis of the matching of the quadratic orders in $\de l_{mn}$ of $a_0$ and $a_1$ with the expansions of the curvature-squared and the curvature-cube Regge action is under way.} of the expansion in $k_{mn}$ of the $a_n(\bar{u}(k_{mn}))$ to be confirmed or rejected.
\end{itemize}
These aspects clearly deserve further study. On the other hand, the analysis we have presented points out a difficulty in the matching of area-Regge-calculus variables with the genuine Loop Quantum Gravity degrees of freedom. The point is the following: the framework of perturbative-area-Regge calculus of section \ref{sec:intro regge calculus} allows to compute correlations of fluctuations of every geometrical object belonging to the $4$-simplex $\Sigma_{12345}$, and not only correlations of area-fluctuations. For instance, correlations of fluctuations of tetrahedron-volumes or of angle-between-triangles are straightforward to compute as such objects can be written in terms of fluctuations of areas (see for instance appendix \ref{app:regge}). The same could be done on the Loop Quantum Gravity side, introducing appropriate functions of the spins in order to represent volumes and angles. Clearly, the correlations computed in this non-standard way will match with the ones computed in perturbative Regge-calculus by construction. The problem comes when one realizes that in Loop Quantum Gravity the volume operator \cite{Rovelli:1994ge,DePietri:1996pj}\cite{Ashtekar:1997fb} and the angle operator \cite{Major:1999mc} can be introduced at the kinematical level and their eigenvalues depend on the intertwiners at the nodes, and not simply on the spins on the links. As the Barrett-Crane model gives trivial dynamics to intertwiner space, we have that correlations of intertwiners will be unrelated to correlations of volumes and angles computed in perturbative quantum Regge-calculus, a fact already noticed in \cite{Bianchi:2006uf} and discussed in great detail in \cite{Alesci:2007tx}. A possible reaction to this fact is to blame the Barrett-Crane model and to  abandon it. Recently, some new models for the dynamics have been proposed \cite{Engle:2007uq,Engle:2007qf,Livine:2007vk,Livine:2007ya,Freidel:2007py,Alexandrov:2007pq}, all of them sharing the property of giving non-trivial dynamics to intertwiner space. From this point of view, a problem which urges to be addressed is to repeat our analysis for the new models, using a semiclassical state $\Psi^{4\textrm{-sim}}_{\Gamma_5,q}(l_{mn},i_n)$ given by a superposition both on spins and on intertwiners with the property of being peaked exactly on the intrinsic and the extrinsic $3$-geometry of the boundary of a regular flat $4$-simplex, instead of the naive state $\Psi_{l_0,\phi_0}(l_{mn})$ of equation (\ref{eq:gaussian x phase}). Luckily, a state of this kind can be built thanks to the result \cite{Rovelli:2006fw}, (see for instance \cite{Alesci:2007tx}). 

Computing correlations for the new models and comparing them with perturbative-Regge-calculus provides a non-trivial test of their viability as models for quantum gravity. Moreover it suggests a new way of interpreting the Barrett-Crane model $W_{\textrm{BC}}(l_{mn})$: namely to understand it as a `partially integrated' model. Suppose for instance that a new model $W_{\textrm{new}}(l_{mn},i_n)$ exists with the following property: the sum over intertwiners on the state $\Psi^{4\textrm{-sim}}_{\Gamma_5,q}(l_{mn},i_n)$ satisfies
\be
\sum_{i_n}\, W_{\textrm{new}}(l_{mn},i_n)\, \Psi^{4\textrm{-sim}}_{\Gamma_5,q}(l_{mn},i_n) = W_{\textrm{BC}}(l_{mn})\, \Psi_{l_0,\phi_0}(l_{mn})\;.
\ee

\vspace{-.8em}

\noindent When using such a model to compute only spin correlations the agreement with perturbative-Regge-calculus is guaranteed by our analysis of the Barrett-Crane model. On the other hand, when computing correlations of a function of the intertwiners such as the volume, the new model should satisfy the non-trivial requirement 
\ben
\sum_{i_n}\, W_{\textrm{new}}(l_{mn},i_n)\,V_1(l_{1m},i_1)V_2(l_{2m},i_2)\, \Psi^{4\textrm{-sim}}_{\Gamma_5,q}(l_{mn},i_n) = W_{\textrm{BC}}(l_{mn}) \,\bar{V}_1(l_{mn})\bar{V}_2(l_{mn})\, \Psi_{l_0,\phi_0}(l_{mn})
\een

\vspace{-.8em}

\noindent with the non-standard expression $\bar{V}_1(l_{mn})$ for the volume in terms of ten spins, an expression  which can be checked against the area-Regge-calculus expression for the volume of a tetrahedron in terms of the ten areas of the $4$-simplex it belongs to. While finding a model $W_{\textrm{new}}(l_{mn},i_n)$ which matches in this way with the Barrett-Crane model could be interesting, this is not at all a requirement for the new models \cite{Engle:2007uq,Engle:2007qf,Livine:2007vk,Livine:2007ya,Freidel:2007py,Alexandrov:2007pq}. The weaker requirement they should satisfy is to have a partially integrated version with properties analogous to the ones postulated above for the Barrett-Crane model and such that it admits an effective description of the form (\ref{eq:Wv BC asymptotics}). This amounts to changing the dynamics in the non-perturbative regime, while leaving the semiclassical description unchanged in form.

\vspace{1em}

\noindent There are three major open issues that the present work does not address:
\begin{itemize}
\item the problem of going beyond the vertex amplitude level, namely the study of composition (a problem addressed at an exploratory level in \cite{Bianchi:2006uf});
\item the problem of the sum over two-complexes (see \cite{Rovelli:2004tv} for a discussion), which corresponds to a sum over the connectivities on the Regge calculus side;
\item the relation with (continuum) perturbative quantum gravity on flat space, possibly viewed as an effective field theory description \cite{Donoghue:1994dn,Burgess:2003jk}.
\end{itemize} 
We hope that the analysis we have presented here could provide a safe starting point for the study of such problems.

\section*{Acknowledgments}
It is a pleasure to thank Carlo Rovelli for continuous help and advice. We are grateful to Pietro Menotti for many illuminating discussions.

\appendix

\section{\texorpdfstring{The geometry of a $4$-simplex: length and area variables}{The geometry of a 4-simplex: length and area variables}}\label{app:regge}
In this appendix we collect some formulas describing the geometry of a $4$-simplex and determine the Jacobian of the transformation from variation of length-variables to variation of area-variables for the regular configuration.

A $4$-simplex is the convex hull of a set of $5$ affinely independent points $\{v_1\mdots v_5\}$ in a Euclidean space of dimension $4$. Subsets of $\{v_1\mdots v_5\}$ define $10$ edges $e_{ij}$ of length $L_{ij}$, $10$ triangles $t_{ijh}$ of area $A_{ijh}$ and $5$ tetrahedra $T_{ijhk}$ of volume $V_{ijhk}$. We call $S_{ijhkl}$ the $4$-volume of the $4$-simplex. Other geometrical objects of interest here are the following: 
\begin{itemize}
\item[-] two triangles $t_{ijh}$ and $t_{ijk}$ sharing the edge $e_{ij}$ identify an angle $\varphi_{(ij)hk}$ which can be defined as $\pi$ minus the angle between the normals to the two triangles within the tetrahedron $T_{ijhk}$;
\item[-] two tetrahedra $T_{ijhk}$ and $T_{ijhl}$ sharing the triangle $t_{ijh}$ identify a dihedral angle $\theta_{(ijh)kl}$ which can be defined as $\pi$ minus the angle between the normals to the two tetrahedra within the $4$-simplex $\Sigma_{ijhkl}$.   
\end{itemize}
The geometry of a $4$-simplex can be completely described giving the lengths $L_{ij}$ of its ten edges, provided that they satisfy a set of triangular inequalities and their higher dimensional analogues. Following \cite{Wheeler:1964rgt}, we introduce the $(n+2)\times (n+2)$ matrix 
{\small
\begin{displaymath}
M_{(n)}(L_{12}\mdots L_{n,n+1})=\left(
\begin{array}{ccccccc}
0 & 1 & 1 & 1&\ldots & \ldots & 1\\
1 & 0 & L_{12}^{\,2} & L_{13}^{\,2} &\ldots &\ldots & L_{1n}^{\,2} \\
1 & L_{12}^{\,2} & 0 & L_{23}^{\,2} & \ldots & \ldots & L_{2n}^{\,2}\\
1 & L_{13}^{\,2} & L_{23}^{\,2} & 0 & \ldots &  \ldots &L_{3n}^{\,2}\\
\vdots & \vdots & \vdots & \vdots & \ddots &  & \vdots \\
1& L_{1,n}^{\,2} & \cdots & \cdots &\cdots & 0 & L_{n,n+1}^{\,2}\\
1& L_{1,n+1}^{\,2} \!\! & L_{2,n+1}^{\,2} \!\! & L_{3,n+1}^{\,2} \!\! &\ldots  & L_{n,n+1}^{\,2}\!\!\!\! & 0
\end{array}\right).
\end{displaymath}
}
The area of the triangle $t_{123}$ can be expressed in terms of the edge lengths $L_{12},L_{13},L_{23}$ as
\be
A_{123}=\frac{1}{4}\sqrt{-\det M_{2}(L_{12},L_{13},L_{23})}\;,\label{eq:Heron}
\ee
an expression which corresponds to the well-known formula of Heron. Similarly, the volume of the tetrahedron $T_{1234}$ can be written as
\be
V_{1234}=\frac{1}{3!\, 2^{\frac{3}{2}}}\sqrt{+\det M_{3}(L_{12},L_{13},L_{14},L_{23},L_{24},L_{34})}\,,\label{eq:Tartaglia}
\ee
which corresponds to Tartaglia's formula. Finally, for the $4$-volume of the $4$-simplex we have
\be
S_{12345}=\frac{1}{4!\, 2^{\frac{4}{2}}}\sqrt{-\det M_{4}(L_{12},L_{13},L_{14},L_{15},L_{23},L_{24},L_{25},L_{34},L_{35},L_{45})}\;.\label{eq:4Tartaglia}
\ee
Triangular, tetrahedral and pentachoral inequalities on the edge lengths correspond to the requirement that the areas of all the ten triangles, the volumes of the five tetrahedra and the $4$-volume of the $4$-simplex are given by positive real quantities when expressed in terms of the edge lengths using formulas (\ref{eq:Heron}), (\ref{eq:Tartaglia}) and (\ref{eq:4Tartaglia}). Angles and dihedral angles can be expressed in terms of areas, volumes and the $4$-volume. The angle $\varphi_{(12)34}$ between the triangles $t_{123}$ and $t_{124}$ can be written in terms of the lengths of the six edges of the tetrahedron $T_{1234}$ they belong to using the following formula
\be
\varphi_{(12)34}(L_{12},L_{13},L_{14},L_{23},L_{24},L_{34}) = \sin^{-1}\frac{3}{2}\frac{L_{12}}{A_{123} A_{124}} V_{1234}\;.
\ee 
Similarly, the dihedral angle $\theta_{(123)45}$ between the two tetrahedra $T_{1234}$ and $T_{1235}$ can be written in terms of the lengths of the ten edges of the $4$-simplex and is given by
\be
\theta_{(123)45}(L_{ij})= \sin^{-1} \frac{4}{3} \frac{A_{123}}{V_{1234} V_{1235}} S_{12345}\;.
\ee
As within a $4$-simplex for each triangle $t_{ijh}$ there is only one dihedral angle $\theta_{(ijh)kl}$, we use the simpler notation $\theta_{(ijh)}$ to identify it. A geometrical property the dihedral angles of a $4$-simplex satisfy is the following
\be
\sum_{1 \leq h<k<l \leq 5}\!\!\! A_{hkl}(L_{mn})\; \frac{\p \theta_{(hkl)}}{\p L_{ij}}(L_{mn})=0\;.\label{eq:Schl id}
\ee
It goes under the name of Schl\"afli differential identity (see the appendix of \cite{Regge:1961px} for a geometrical proof). 
 
We call regular a $4$-simplex which has all its $10$ edge-lengths equal. Using Bang's theorem, Baez et al. \cite{Baez:2002rx} show that a non-degenerate $4$-simplex which has all its $10$ triangle-areas equal is necessarily regular. Now we consider fluctuations of edge-lengths $\de L_{mn}$ around the regular configuration with lengths $L_0$ and show that the nearly-regular $4$-simplex they define can be described in terms of triangle-area fluctuations $\de A_{hkl}$ around the regular configuration with areas $A_0=\frac{\sqrt{3}}{4}L_0^2$. In fact the Jacobian matrix $U$ from length-fluctuations to area-fluctuations (see equation (\ref{eq:U-V})) can be computed using formula (\ref{eq:Heron}) and turns out to be invertible. We call $V$ its inverse. Choosing the following ordering for the fluctuation vectors
\begin{align*}
\{\de L_{mn}\}=&\,(\de L_{12},\de L_{13},\de L_{14},\de L_{15},\de L_{23},\de L_{24},\de L_{25},\de L_{34},\de L_{35},\de L_{45})\;,\\
\{\de A_{ijh}\}=&\,(\de A_{345},\de A_{245},\de A_{235},\de A_{234},\de A_{145},\de A_{135},\de A_{134},\de A_{125},\de A_{124},\de A_{123})\;,
\end{align*}
the matrices $U$ and $V$ turn out to be symmetric and have the following coefficients\footnote{The matrices (\ref{eq:Umatrix}),(\ref{eq:Vmatrix}) and (\ref{eq:K LQG}),(\ref{eq:dS/dAdA}) all have the same structure which corresponds to the following matrix
\begin{equation*}
K(a,b,c)=\left(
\begin{array}{cccccccccc}
a & b & b & b & b & b & b & c & c & c \\
b & a & b & b & b & c & c & b & b & c \\ 
b & b & a & b & c & b & c & b & c & b \\ 
b & b & b & a & c & c & b & c & b & b \\ 
b & b & c & c & a & b & b & b & b & c \\ 
b & c & b & c & b & a & b & b & c & b \\ 
b & c & c & b & b & b & a & c & b & b \\ 
c & b & b & c & b & b & c & a & b & b \\ 
c & b & c & b & b & c & b & b & a & b \\
c & c & b & b & c & b & b & b & b & a
\end{array}\right)
\end{equation*}
with specific coeffeicients $a,b,c$. Here we collect some useful formulae for the eigenvalues of a matrix of this form: 
\begin{align*}
\lambda_0=&\, a+6 b +3 c \qquad \textrm{(non-deg.)}\\
\lambda_1=&\, a+ b -2 c \qquad \textrm{(4-fold deg.)}\\
\lambda_2=&\, a-2 b + c \qquad \textrm{(5-fold deg.)}
\end{align*}
The eigenvalue $\lambda_0$ is associated to the eigenvector $v_0=(1,1,1,1,1,1,1,1,1,1)$.
}:
{\small
\begin{equation}
U=\frac{L}{2\sqrt{3}}\left(
\begin{array}{cccccccccc}
0 & 0 & 0 & 0 & 0 & 0 & 0 & 1 & 1 & 1 \\
0 & 0 & 0 & 0 & 0 & 1 & 1 & 0 & 0 & 1 \\
0 & 0 & 0 & 0 & 1 & 0 & 1 & 0 & 1 & 0 \\
0 & 0 & 0 & 0 & 1 & 1 & 0 & 1 & 0 & 0 \\
0 & 0 & 1 & 1 & 0 & 0 & 0 & 0 & 0 & 1 \\
0 & 1 & 0 & 1 & 0 & 0 & 0 & 0 & 1 & 0 \\
0 & 1 & 1 & 0 & 0 & 0 & 0 & 1 & 0 & 0 \\
1 & 0 & 0 & 1 & 0 & 0 & 1 & 0 & 0 & 0 \\
1 & 0 & 1 & 0 & 0 & 1 & 0 & 0 & 0 & 0 \\
1 & 1 & 0 & 0 & 1 & 0 & 0 & 0 & 0 & 0
\end{array}\right)\;,\label{eq:Umatrix}
\end{equation}
}
and
{\footnotesize
\begin{equation}
V=\frac{1}{\sqrt{3} L}{ \left(
\begin{array}{cccccccccc}
2 & -1 & -1 & -1 & -1 & -1 & -1 & 2 & 2 & 2 \\
-1 & 2 & -1 & -1 & -1 & 2 & 2 & -1 & -1 & 2 \\
-1 & -1 & 2 & -1 & 2 & -1 & 2 & -1 & 2 & -1 \\
-1 & -1 & -1 & 2 & 2 & 2 & -1 & 2 & -1 & -1 \\
-1 & -1 & 2 & 2 & 2 & -1 & -1 & -1 & -1 & 2 \\ 
-1 & 2 & -1 & 2 & -1 & 2 & -1 & -1 & 2 & -1 \\ 
-1 & 2 & 2 & -1 & -1 & -1 & 2 & 2 & -1 & -1 \\
2 & -1 & -1 & 2 & -1 & -1 & 2 & 2 & -1 & -1 \\
2 & -1 & 2 & -1 & -1 & 2 & -1 & -1 & 2 & -1 \\
2 & 2 & -1 & -1 & 2 & -1 & -1 & -1 & -1 & 2 
\end{array}\right)}\;.\label{eq:Vmatrix}
\end{equation}
}
Using the matrix (\ref{eq:Vmatrix}), the geometry of a nearly-regular $4$-simplex can be completely described in terms of area-fluctuations: for instance the fluctuation of the volume of the tetrahedron $T_{1234}$ is given by equation
\be
\de V_{1234}=\sum_{1\leq h < k < l \leq 5}\;\Big(\;\sum_{1\leq i < j \leq 4}\frac{\p V_{1234}}{\p L_{ij}}(L_0)\;\;V_{(ij)(hkl)}\Big)\;\de A_{hkl}
\ee 
in terms of the ten area-fluctuations.




\end{document}